\documentclass[10pt]{iopart}

\usepackage{iopams}
\usepackage{rsfs}
\usepackage{setstack}
\usepackage[all]{xy}
\usepackage{graphicx}
\newcommand{\be}{\begin{equation}}
\newcommand{\ee}{\end{equation}}
\newcommand{\bea}{\begin{eqnarray}}
\newcommand{\eea}{\end{eqnarray}}

\newcommand{\ra}{\rangle}
\newcommand{\pa}{\partial}
\newcommand{\la}{\langle}

\newcommand{\La}{\la\la}
\newcommand{\Ra}{\ra\ra}
\begin{document}

\title[Work Required for Selective Quantum Measurement]{Work Required for Selective Quantum Measurement}

\author{Eiji Konishi}

\address{Graduate School of Human and Environmental Studies, Kyoto University, Kyoto 606-8501, Japan}
\ead{konishi.eiji.27c@kyoto-u.jp}
\vspace{10pt}
\begin{indented}
\item[]\today
\end{indented}

\begin{abstract}
In quantum mechanics, we define the measuring system $M$ in a selective measurement by two conditions.
Firstly, when we define the measured system $S$ as the system in which the non-selective measurement part acts, $M$ is independent from the measured system $S$ as a quantum system in the sense that any time-dependent process in the total system $S+M$ is divisible into parts for $S$ and $M$.
Secondly, when we can separate $S$ and $M$ from each other without changing the unitary equivalence class of the state of $S$ from that obtained by the partial trace of $M$, the eigenstate selection in the selective measurement cannot be realized.
In order for such a system $M$ to exist, we show that in one selective measurement of an observable of a quantum system $S_0$ of particles in $S$, there exists a negative entropy transfer from $M$ to $S$ that can be directly transformed into an amount of Helmholtz free energy of $k_BT$ where $T$ is the thermodynamic temperature of the system $S$.
Equivalently, an extra amount of work, $k_BT$, is required to be done by the system $M$.
\end{abstract}

%
%
%
%
%

\section{Introduction}

It is a fundamental question whether quantum measurement in itself is a physical process with energy transfer.
The thermodynamics of information\cite{Landauer,Ben,LandauerPT,Shizume,Piechocinska,SU1,SU2,SU3,SU4,SagawaPTP,DJ,Sagawa} asserts that information processing assuming the agent's memory is a physical process.
As the most elementary example, in the feedback process of a classical measurement in the Szil${\acute{{\rm a}}}$rd engine, to erase one bit of information in a symmetric potential memory of the feedback agent requires an amount of work of $\ln 2$ times $k_BT$\cite{Landauer,Shizume,Piechocinska,SU2,Szilard,Brillouin}.\footnote{In this paper, $T$ denotes the thermodynamic temperature of the system.
Here, {\it the system} is the memory and the heat bath\cite{SU2}.
In the main statement, {\it the system} is the measured system.}
However, the thermodynamics of information simply uses measurement as one step in its protocols and the fundamental question is still unaddressed.

The difficulty of formulating this question as a problem in physics comes from the fact that, in the Copenhagen interpretation of quantum mechanics\cite{Born,Bohr}, the definition of the {\it measuring system} that completes selective measurements is unclear.
The infamous {measurement problem} arises from the fact that there is no clear distinction between the measured system and the measuring system as their combined system should obey the causal, continuous and reversible change expressed in the Schr$\ddot{{\rm o}}$dinger equation.

Regarding this aspect of the Copenhagen interpretation, a long time ago, von Neumann noticed the logical consequence that there arises an infinite regression of measuring systems in the selective measurement process in the framework of causal, continuous and reversible (i.e., unitary) changes and he introduced the projection hypothesis to complete this infinite regression\cite{Neumann,Luders}.
In further studies after von Neumann, instead of selective measurement, {\it non-selective measurement}\cite{NC} was introduced and considered as the prototype of measurements in the program of decoherence\cite{Decoherence,Decoherence2a,JoosZeh,Decoherence2a2,Decoherence2b,Decoherence2c}.
Non-selective measurement describes the measurement result statistically in an ensemble of copies of a quantum system in terms of the density matrix and its reduction is weaker than that of selective measurement.
The non-selective measurement process does not require the projection hypothesis.

Here, we give clear-cut definitions of {\it non-selective} and {\it selective measurements} of an observable with a discrete spectrum.
When we express the density matrix of the combined system of the measured system and the measurement apparatus in the eigenbasis of this observable and the pointer's variable, the former refers to the vanishing of all off-diagonal elements of the density matrix while no diagonal element changes: the resultant state is a statistical mixture of eigenstates with weights given by the Born rule.
The latter refers to the non-selective measurement plus its subsequent selection process of a diagonal element of the density matrix (i.e., the {\it event reading}): the resultant state is a pure eigenstate.

In this paper, we model the {\it measurement process} by non-selective measurement plus its subsequent event reading, the latter part of which has not been treated as a process in measurement theories.
The stance of this modeling differs from the stance of letting an event be a dynamical concept in the program of decoherence\cite{DecoherenceBook}.
Based on this modeling, we quantify the entropy production on the measured system that accompanies the non-selective measurement due to the event reading in order to answer the question posed at the beginning.\footnote{Throughout this paper, we refer to entropy divided by $k_B$ as {\it entropy}.
This is not a change of dimensions but a change of terminology.
We argue that {\it entropy is produced from selective measurement} by the finite variation of the logarithm of the normalization constant of the density matrix in the quantum system, just as the entropy production in the canonical distribution attributed to the change of the Helmholtz free energy comes from the finite variation of the logarithm of the partition function.
Since this entropy production is independent from the internal energy, {\it absorption of heat does not accompany this entropy production}.
The definition of the {\it entropy transfer} accompanying selective measurement is given by Eq.(\ref{eq:Key0}) in Sec. 2.}
Subsequently, we propose a criterion that sheds light on the concept of a {\it measuring system} that is able to select a quantum eigenstate, primarily of the measuring system, from an exclusive mixture.

When a measuring system that satisfies our criterion
(specifically, the {\it type I} system defined in Secs. 2 and 4)
 exists, the main statement of this paper is as follows.

\smallskip
\smallskip

{\it{In a quantum system of particles, a selective measurement of an observable requires negative entropy production $\sigma=-1$ on the system that can be directly transformed into a positive amount of Helmholtz free energy of $-\sigma k_BT=k_BT$.
Equivalently, it requires an amount of work of $k_BT$ done by the measuring system.}}

\smallskip
\smallskip

Now, we set up the measurement process used in the discussion of this paper.

Firstly, we assume a quantum system of particles and a given discrete or discretized continuous observable of this quantum system.

Secondly, we do {one} non-selective measurement of this observable, which we temporally contract to an instantaneous non-selective measurement in order to facilitate the analysis and clarify the quantification of the effect of one measurement.

Then, after an entangling interaction, the situation {\it just before} the event reading is realized and we are able to study the role of the measuring system in the eigenstate selection (i.e., the event reading) as the sequel of non-selective measurement.

In this paper, we assume, as a concrete model of {non-selective} measurement process, a three-tier preparation of the systems: the measured system, a macroscopic measurement apparatus and the measuring system, and we use the mechanism of a continuous superselection rule\cite{Araki}.

We use the mechanism of a continuous superselection rule as the model of non-selective measurement because this mechanism is a closed process governed by a Hamiltonian (see the explanations in Appendix D) and is thus compatible with a setting where the quantum version of the Jarzynski equality\cite{Jarzynski,Tasaki1,Mukamel,ReviewJ,Experiment} (this equality will be invoked in Sec. 3) holds.
In contrast, the usual decoherence mechanism arising from the interactions with the environmental system is explained as the tracing out of the degrees of freedom of the environmental system and is not a closed process of the target system governed by a Hamiltonian.

Here, we explain the basic concepts with respect to non-selective measurement for comprehensiveness.

In the context of non-selective measurement, quantum mixed states are used.
A {\it quantum mixed state} is given by the density matrix where, in a statistical ensemble of copies of a quantum system, due to our lack of knowledge about the system, the statistical probability that this state is realized as the pure state $|\Psi\ra$ satisfying $\la \Psi|\Psi\ra=1$ is given by $w$
\begin{eqnarray}
\widehat{\varrho}&=&\sum_n w_n |\Psi_n\rangle\langle \Psi_n|\;,\label{eq:mix}\\
1&=&\sum_n w_n\;.\label{eq:p}
\end{eqnarray}
The statistical fluctuation in this mixed state is independent from the quantum fluctuation arising from the quantum superposition in pure states.

In the density matrix, for example, for the projection operator ${\widehat{P}(x)}$ of a discrete observable's value $x$
\begin{equation}
\widehat{\varrho}\to \frac{{\widehat{P}(x)}\widehat{\varrho} {\widehat{P}(x)}}{{\mbox{tr}}({\widehat{P}(x)}\widehat{\varrho})}
\end{equation}
is the state reduction used in the selective measurement as a stochastic process\cite{Neumann,Luders}.
In addition to this, non-selective measurement such that
\begin{eqnarray}
\widehat{\varrho}&\to&\sum_{{\rm all}\ y}{\widehat{P}(y)}\widehat{\varrho}{\widehat{P}(y)}\;,\label{eq:P0}\\ 
\widehat{1}&=&\sum_{{\rm all}\ y}\widehat{P}(y)\label{eq:P}
\end{eqnarray}
is possible for a statistical ensemble of copies of a quantum system\cite{DRM}.
By Eq.(\ref{eq:P0}), the off-diagonal elements of the density matrix vanish.
In non-selective measurement, we do not consider a particular quantum state resulting from measurement but consider the probability of measurement results, that is, the statistical result for an ensemble of many copies of a quantum system.
So, non-selective measurement does {\it not} contain the eigenstate selection process and the mixed state of the total system for measurement obtained by this non-selective measurement is the state {\it just before} the event reading.
We stress that the density matrix of this mixed state does not say anything about the event realized by the quantum eigenstate selection but only refers to the probability of realization of each event.

To describe the measurement process, besides the eigenstate $|x\ra$ of a measured observable $\widehat{{\mathscr{O}}}$ of our measured system, we assume the eigenstate $|{\mathfrak M}_x\ra$ of the discrete or finely discretized continuous pointer variable of the measuring system for the measurement result ${\mathfrak M_x}$ corresponding to $x$, and the initial state $|{\mathfrak A}_0\ra$ of the measurement apparatus.
We denote the eigenstate $|x\ra|{\mathfrak A}_0\ra$ to which a continuous superselection rule has been applied by $|x,{\mathfrak A}_0\ra$ or the double ket $|x\Ra$.

Throughout this paper, we denote operators and superoperators with a hat and a wide tilde, respectively.

The organization of this paper and brief intuitive explanations for the results are as follows.

The next section consists of three parts.
Firstly, we formulate our measurement scheme in the Schr$\ddot{{\rm o}}$dinger picture.
Secondly, we analyze the single instantaneous non-selective measurement process as a cut-off inhomogeneous one-time Poisson process by the density matrix of the total system for measurement in the statistical treatment, where the statistical ensemble is enlarged.
Thirdly, we discuss the entropy transfer from the measuring system to the rest system accompanying the selective measurement process based on the precise definition of a {\it measuring system} that can select its quantum eigenstate from an exclusive mixture.

Here, the entropy transfer indicates the non-divisibility of the state of the total system into those of the subsystems when holding the unitary equivalence classes of the states of the subsystems constant (note footnote $+$), and {\it the transferred entropy originates in the reduction of our knowledge about the total system}.
This reduction is due to the averaging operation in the statistical treatment of a non-selective measurement occurrence as a cut-off one-time Poisson process.
In the total system, this knowledge that is an {\it analogue of} information is lost by the measuring system, and the lost knowledge by the diagonal part of the density matrix is gained by the combined measured system.
For this fact, the work is required to be done by the measuring system to the combined measured system.

In Sec. 3, we consider the thermodynamics of the {\it measured} system and the measurement apparatus in the Heisenberg picture by adopting the direct treatment of non-selective measurement occurrences and incorporate the result of Sec. 2 into the second law of thermodynamics.
At this time, we invoke the quantum version of the Jarzynski equality\cite{Jarzynski,Tasaki1,Mukamel,ReviewJ,Experiment}.
Then, we show the main statement of this paper.

In Sec. 4, we briefly summarize the resultant arguments and compare our results with those of three other theories.

In Appendix A, we explain the grounds for asserting quantum mechanical equivalence between the direct description and the statistical description of non-selective measurement occurrence.
In Appendix B, we mathematically formulate the statement of von Neumann's infinite regression of measuring systems.
In Appendix C, we explain the two energy measurement approach to defining the moment-generating function of quantum work, perform the derivations of two formulae in Sec. 3, and explain via measurement theory recent developments in studies of the definition of quantum work and the quantum Jarzynski equality.
In Appendix D, we give a brief account of the mechanism for non-selective measurement in the combined system of the measured system and the macroscopic measurement apparatus as the consequence of a continuous superselection rule.

As a last overall point in this introduction, we comment on how we describe non-selective measurement occurrences in the main text.
In extant measurement theories (e.g., the program of decoherence\cite{Decoherence,Decoherence2a} and the Ghirardi--Rimini--Weber (GRW) model\cite{GRW}) that treat non-selective measurements, we use {\it either} direct\cite{Decoherence,Decoherence2a} {\it or} statistical\cite{GRW} description of non-selective measurement occurrence.
However, to derive the main statement of this paper, we need to combine results from {\it both} direct {\it and} statistical descriptions of non-selective measurement occurrence, which are complementary.
This is a novel consequence and happens because in this paper we treat {\it event reading} alongside non-selective measurement; {\it event reading is a process lying outside of extant measurement theories}.
At present, we treat event reading by considering its mechanism to be a black box.
Specifically, in the direct description of non-selective measurement occurrence, {\it to read an event by a measuring system} we require non-unitary overall factors in the density matrices of the combined measured system and the measuring system.
This is the consequence of von Neumann's infinite regression of measuring systems.
To derive these factors as the consequence of an {\it entropy transfer}, we need the statistical description of non-selective measurement occurrence.

\section{Measurement Process}
\subsection{Scheme of the measurement}
First, we explain our model of {\it quantum measurement}.
Here, we take a quantum pure state (i.e., with no statistical factors) that obeys the von Neumann equation as the initial state.

We assume three systems.
First is the measured system $S_0$.
Second is the macroscopic measurement apparatus $A$, which is abstracted to a quantum system with one degree of freedom and leads to a non-selective measurement in the combined system $S_0+A$ without any interaction with outer systems due to a continuous superselection rule (for its details, see Appendix D).
As the continuous superselection rule, we consider a macroscopic physical quantity, for example, the center of mass momentum of the {macroscopic} measurement apparatus $A$, which is regarded as a classical observable to a good approximation.
The initial state of the macroscopic measurement apparatus $A$ in the non-selective measurement process is assumed to have an ignorable but finite quantum uncertainty of the continuous superselection rule.
Third is the measuring system $M$ that can read the event after a non-selective measurement.

Here, we must add a note.
In the presence of a continuous superselection rule, there is a no-go theorem proved by Araki in Ref.\cite{Araki2}.
This no-go theorem asserts that, in an infinite time process, the measurement apparatus $A$ used to separate the continuous superselection sectors of $S_0+A$ cannot record the measurement results as in Eq.(\ref{eq:step2}) and thus cannot be used for the event reading.
Thus, we require the second measurement apparatus $M$ for the event reading.
So, this three-tier preparation of the systems is a well-accepted setting for selective measurement.\cite{Araki2}

As stated in Introduction, we denote the eigenstate of a measured observable $\widehat{{\mathscr{O}}}$, of the measured system $S_0$, corresponding to the eigenvalue ${x}_n$ of $\widehat{{\mathscr{O}}}$ by $|{x}_n\ra$, and denote the discrete or finely discretized continuous pointer variable of the measuring system $M$ by ${\mathfrak M}$.

Since we assume a continuous superselection rule in the system $A$, the observable $\widehat{{\mathscr{O}}}\otimes\widehat{1}^A$ is restricted to a direct integral of operators acting in their superselection sectors in the system $S_0+A$ (see Eq.(\ref{eq:B8}) in Appendix D).

\begin{figure}[htbp]
\begin{center}
\includegraphics[width=0.8\hsize,bb=0 0 250 100]{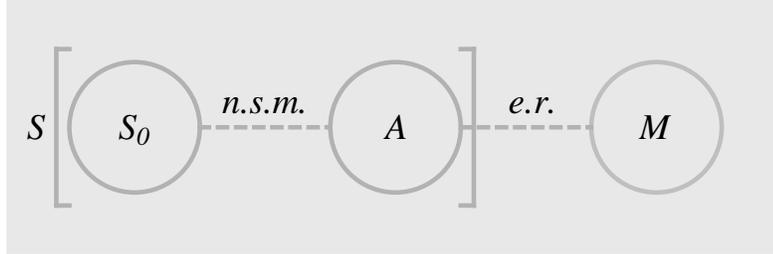}
\end{center}
\caption{
Our scheme for quantum measurement is schematically shown.
The combined measured system $S$, which consists of the quantum system $S_0$ and the macroscopic measurement apparatus (the separation apparatus) $A$, undergoes non-selective measurement (n.s.m.) due to the continuous superselection rule in $A$.
After this non-selective measurement and the entangling interaction between the systems $S$ and $M$, the event reading (e.r.) is done by the measuring system $M$.}
\end{figure}

Our scheme for quantum measurement of $\widehat{{\mathscr{O}}}$ consists of the following four independent steps for the density matrix $\widehat{\varrho}$ of the system $S_0+A+M$ (see Fig.1).
(In equations, the right arrow indicates the change of the density matrix according to the corresponding process.)
\begin{enumerate}
\item The initial statistical ensemble of copies of the system $S_0+A+M$ at time $t=t_{\rm in}$ is a pure ensemble
\begin{equation}
\widehat{\varrho}=\Biggl(\sum_nc_n|{x}_n,{\mathfrak A}_0\ra\Biggr)|{\mathfrak M}_0\ra\la {\mathfrak M}_0|\Biggl(\sum_m \bar{c}_m\la {x}_m,{\mathfrak A}_0|\Biggr)\;.\label{eq:step1}
\end{equation}
Here, we assume the condition $\sum_n|c_n|^2=1$ and the {\it non-triviality of the initial state} such that at least two of $c_n$ are non-zero.
In this step, the system $M$ is in an isolated state.
\item Non-selective measurement of $\widehat{{\mathscr{O}}}$ in the system $S_0+A$ within the time interval $t_{\rm in}\le t\le t_{\rm out}$  changes the pure ensemble to an exclusive mixture:
\begin{eqnarray}
&&\widehat{\varrho}=\Biggl(\sum_nc_n|{x}_n,{\mathfrak A}_0\ra\Biggr)|{\mathfrak M}_0\ra\la {\mathfrak M}_0|\Biggl(\sum_m \bar{c}_m\la {x}_m,{\mathfrak A}_0|\Biggr)\nonumber\\
&&\longrightarrow\sum_n|c_n|^2|{x}_n,{\mathfrak A}_0\ra|{\mathfrak M}_0\ra\la {\mathfrak M}_0|\la {x}_n, {\mathfrak A}_0|\;.\label{eq:step2}
\end{eqnarray}
Here, a continuous superselection rule in the system $A$ is assumed.
\item The system $S_0+M$ causally and continuously changes according to the von Neumann equation until $t=t_0-\epsilon> t_{\rm out}$\footnote{Throughout this paper, we let $\epsilon$ denote a positive infinitesimal time increment.} by energy feedback (when ${\mathfrak M}$ refers to energy) or von Neumann-type entangling interaction (when ${\mathfrak M}$ refers to a pointer coordinate):
\begin{eqnarray}
&&\widehat{\varrho}=\sum_n|c_n|^2|{x}_n,{\mathfrak A}_0\ra|{\mathfrak M}_0\ra\la {\mathfrak M}_0|\la {x}_n, {\mathfrak A}_0|\nonumber\\
&&\longrightarrow\sum_n|c_n|^2|{x}_n,{\mathfrak A}_0\ra|{\mathfrak M}_n\ra\la {\mathfrak M}_n|\la{x}_n,{\mathfrak A}_0|\;.\label{eq:step3}
\end{eqnarray}
\item Reading of the pointer variable ${\mathfrak M}$ of the system $M$ at $t=t_0$ changes the exclusive mixture to a pure ensemble:
\begin{eqnarray}
&&\widehat{\varrho}=\sum_n|c_n|^2|{x}_n,{\mathfrak A}_0\ra|{\mathfrak M}_n\ra\la {\mathfrak M}_n|\la {x}_n,{\mathfrak A}_0|
\nonumber\\
&&\longrightarrow|{x}_{n_0},{\mathfrak A}_{0}\ra|{\mathfrak M}_{n_0}\ra\la {\mathfrak M}_{n_0}|\la {x}_{n_0},{\mathfrak A}_{0}|\;,\label{eq:step4}
\end{eqnarray}
which is a stochastic event acausally occurring with the probability $|c_{n_0}|^2$ according to the Born rule.
In this step, it is assumed that the system $M$ is reset to an isolated state (see also footnote $+$).
\end{enumerate}
In this scheme, {\it non-selective measurements} of $S$ and of $S+M$ refer to steps (i) and (ii) for $S$ and to steps (i), (ii) and (iii) for $S+M$, and {\it selective measurement} refers to steps (i), (ii), (iii) and (iv).

In the following, we denote the composite system $S_0+A$ by ${{S}}$.
In Secs. 2 and 3, we refer to $S_0$ and $S$ as the {\it measured system} and the {\it combined measured system}, respectively, since the heart of a measurement in the scheme lies in step (iv).

\subsection{Non-selective measurement process}
In this subsection, we study the time evolution of the total system during $t_{\rm in}\le t<t_0$.

We denote the total time-dependent {\it generalized Hamiltonian} (which needs to be generalized to contain an arbitrary Hermitian operator when ${\mathfrak M}$ refers to the energy of the system $M$; in the following, we refer to this as the {\it Hamiltonian}) of the total system $S_0+A+M$ including the interactions by $\widehat{{\cal H}}_{{\rm tot}}(t)$ and denote the sum of the kinetic Hamiltonians $\widehat{{\cal H}}_{\rm kin}^{S_0}\otimes \widehat{1}^A\otimes \widehat{1}^M$, $\widehat{1}^{S_0}\otimes \widehat{{\cal H}}^A_{{\rm kin}}\otimes \widehat{1}^M$ and $\widehat{1}^{S_0}\otimes \widehat{1}^A\otimes \widehat{{\cal H}}^M_{{\rm kin}}$ by $\widehat{{\cal H}}_{\rm kin}$.
Here, the time dependence of $\widehat{{\cal H}}_{{\rm tot}}(t)$ reflects the protocol of measurement.

The Hamiltonian $\widehat{{\cal H}}_{{\rm tot}}(t)-\widehat{{\cal H}}_{\rm kin}$ for the interactions between the systems $S_0$ and $A$ and between the systems $S_0$ and $M$ (under feedback control by an external agent in the case where ${\mathfrak M}$ refers to the energy of the system $M$) is the interaction Hamiltonian $\widehat{{\cal H}}_{{\rm int}}(t)$, that is, the Hermitian operator\cite{Neumann}
\begin{eqnarray}
\widehat{{\cal H}}_{\rm tot}(t)&=&\widehat{{\cal H}}_{\rm kin}+\widehat{{\cal H}}_{\rm int}(t)\;,\\
&&\nonumber\\
\widehat{{\cal H}}_{\rm int}(t)&=&\left\{\begin{array}{cc}0&t= t_{\rm in}\\
&\\
-(\Lambda^A\cdot 1^A)\widehat{{\mathscr{O}}}^{S_0}\otimes \widehat{P}^A\otimes \widehat{1}^M& t_{\rm in}< t\le  t_{\rm out}\\
&\\
\widehat{{\cal H}}_{\rm fb}(\widehat{{\mathscr{O}}}^{S_0}\otimes \widehat{1}^A\otimes \widehat{1}^M)&t_{\rm out}< t<t_0
\end{array}\right.
\end{eqnarray}
for the center of mass momentum operator $\widehat{P}^A$ of the system $A$ as the continuous superselection rule in the system $A$ and the {\it Hermitian operator} $\widehat{{\cal H}}_{\rm fb}$.
The Hermitian operator $\widehat{{\cal H}}_{\rm fb}$ represents two distinct cases.
In the first case, ${{\mathfrak M}}$ refers to the energy of the system $M$.
Here, $\widehat{{\cal H}}_{\rm fb}$ gives rise to the generator of an energy feedback unitary transformation (i.e., Eq.(\ref{eq:step3})), with its strength
\begin{equation}
\Lambda^M\cdot 1^M \equiv \frac{{\mathfrak M}_n-{\mathfrak M}_0}{x_n(t_0-t_{\rm out})}
\end{equation}
designed to be common to all $n$, in the open quantum system $S+M$ by tracing out the energy reservoir.
In the second case, ${{\mathfrak M}}$ refers to a pointer coordinate.
Here, $\widehat{{\cal H}}_{\rm fb}$ is a von Neumann-type interaction Hamiltonian $-(\Lambda^M\cdot 1^M)\widehat{{\mathscr{O}}}^{S_0}\otimes \widehat{1}^A\otimes \widehat{{\mathfrak P}}^M_c$ for the canonically conjugate operator $\widehat{{\mathfrak P}}^M_c$ of the continuous pointer position operator $\widehat{{\mathfrak M}}^M_c$ of the measuring system $M$ such that $[\widehat{{\cal H}}^M_{{\rm kin}},\widehat{{\mathfrak P}}^M_c]=0$ holds.
In $\widehat{{\cal H}}_{\rm int}(t)$, both $\Lambda^A$ and $\Lambda^M$ are dimensionless positive-valued constants, while unities $1^A$ and $1^M$ have dimensions.

It is assumed that $\Lambda^A$ and $\Lambda^M$ in $\widehat{{\cal H}}_{\rm int}(t)$ are strong enough that we can neglect $\widehat{{\cal H}}_{\rm kin}$ during steps (ii) and (iii) by using time parameters rescaled by the factors $\Lambda^A$ (for step (ii)) and $\Lambda^M$ (for step (iii)) as $\delta t_{\rm old} \to \delta t_{\rm new}=\Lambda \delta t_{\rm old}$ ($\Lambda=\Lambda^A,\Lambda^M$), respectively, due to the large effective masses in the kinetic part of the rescaled von Neumann equation.
From this assumption, it is sufficient for the time intervals $t_{\rm out}-t_{\rm in}$ and $t_0-t_{\rm out}$ for steps (ii) and (iii), respectively, to be {\it short} for the original time parameter $t$ in order for these steps to be Eqs.(\ref{eq:step2}) and (\ref{eq:step3}), respectively.
However, the time interval $t_{\rm out}-t_{\rm in}$ for step (ii) must be long enough for the time parameter rescaled by $\Lambda^A$ that the mechanism of the continuous superselection rule works.

The von Neumann equation for the density matrix of the total system before the event reading (i.e., $t<t_0$) is
\begin{eqnarray}
\widehat{\varrho}(t+dt)=\widehat{\varrho}(t)-\frac{i}{\hbar}[\widehat{{\cal H}}_{\rm kin}+\widehat{{\cal H}}_{\rm int}(t),\widehat{\varrho}(t)]dt\;.
\end{eqnarray}

Here, we clarify the idea of {\it neglecting the kinetic Hamiltonian} in the context of a quantum measurement.
As an example, we consider a quantum measurement of the position of a particle.
Because the position operator has a continuous spectrum, the norms of its eigenvectors diverge, and thus none of its eigenvectors is a state vector.
So, as mentioned in the Introduction, we need to adopt discretized position eigenvalues, such as the compartments of a partitioned box.
We note that the eigenvalues of this discretized position in the superposition of the measured particle need to be definite during the measurement process.
We can neglect the kinetic Hamiltonian during the measurement process if it does not change these eigenvalues.

\begin{figure}[htbp]
\begin{center}
\includegraphics[width=0.5\hsize,bb=0 0 260 290]{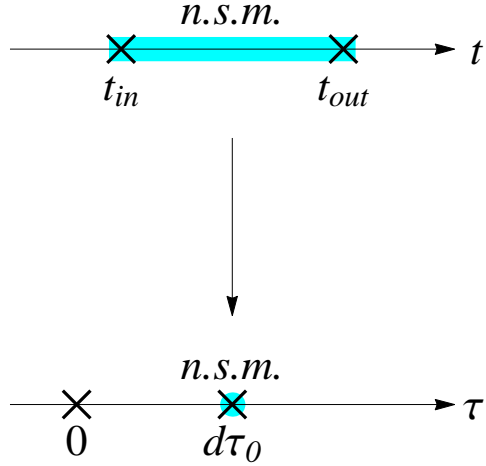}
\end{center}
\caption{
We contract the time interval $I\equiv [t_{\rm in},t_{\rm out}]$ in the time-axis $t$ to an instance $\tau=d\tau_0$ (a point set in ${\bf R}$ is a closed set) in the new time-axis $\tau$.
Within this time interval, the non-selective measurement (n.s.m.) of the system $S$ occurs.
Since unitary change driven by a Hamiltonian is a continuous change, an infinitesimal unitary change such as the change during $0\le \tau<d\tau_0$ in the time-axis $\tau$ does not change the state.}
\end{figure}

In the rest of this section, to facilitate the analysis of the measurement process, we contract the non-selective measurement process of the system $S$ driven by $\widehat{{\cal H}}_{\rm int}$ during the time interval $I\equiv [t_{\rm in},t_{\rm out}]$ to an {\it instantaneous} non-selective measurement event at $\tau=d\tau_0$ ($d\tau_0$ is a positive infinitesimal time increment) with a new time parameter (see Fig.2)
\begin{equation}
\tau(t)=\left\{\begin{array}{cc}
t-t_{\rm in}+d\tau_0&t<t_{\rm in}\\
&\\
d\tau_0&t_{\rm in}\le  t\le  t_{\rm out}\\
&\\
t-t_{\rm out}+d\tau_0&t_{\rm out}< t\;.
\end{array}\right.\label{eq:TAUT}
\end{equation}
This practical approach of contracting the time interval is possible because we can solve the time evolution within $I$ by
\begin{equation}
\widehat{\varrho}(t_{{\rm out}})=\sum_{{\rm all}\ y} \widehat{P}(y)\widehat{\varrho}(t_{{\rm in}})\widehat{P}(y)\;.\label{eq:Iprocess}
\end{equation}
Here, $\widehat{P}(y)$ is a projection operator\footnote{In the notation of Appendix D, this operator can be rewritten as $(\int^\bigoplus |y(p)\ra\la y(p)| dp)\otimes \widehat{1}^M$, where we set $|y(p)\ra\equiv|y\ra$.}
\begin{equation}
\widehat{P}(y)\equiv |y\ra\la y|\otimes \widehat{1}^A \otimes \widehat{1}^M\;.
\end{equation}
Of course, by this contraction, the information about the time evolution during step (ii) is completely lost.
However, this information is unnecessary for our analysis in this section.

\begin{figure}[htbp]
\begin{center}
\includegraphics[width=0.5\hsize,bb=0 0 260 230]{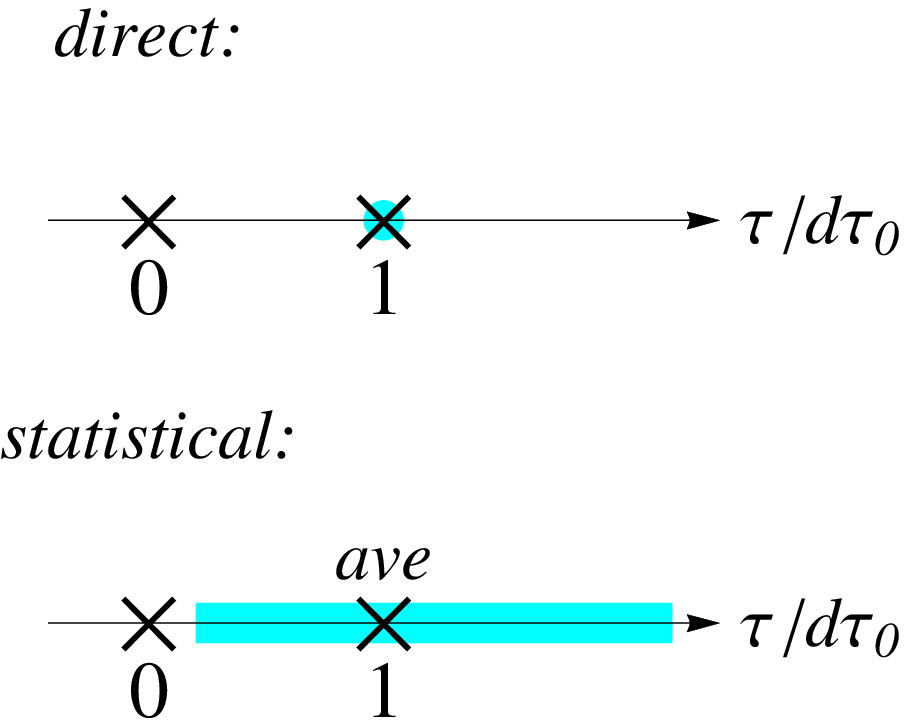}
\end{center}
\caption{
Two descriptions of non-selective measurement occurrence are schematically shown after the temporal contraction of $I$.
In the direct description (upper panel), the instance $\tau=d\tau_0$ in Fig.2 is definite.
In contrast, in the statistical description (lower panel), every possible instance corresponding to $\tau=d\tau_0$ in Fig.2 is an instance of a stochastic event of a one-time Poisson process that starts from $\tau=0$ with characteristic time $d\tau_0$ (i.e., the averaged occurrence time (ave)) and occurs in the enlarged ensemble.
Here, note that only one event occurs in the individual ensemble.
Since every stochastic event $\tau=\epsilon$ in this one-time Poisson process refers to an occurrence time of the same non-selective measurement process (\ref{eq:Iprocess}) in the new time-axis $\tau$, this one-time Poisson process in the new time-axis $\tau$ is separated from the process (\ref{eq:Iprocess}).}
\end{figure}

From here and for a while, beginning with the above setup, we drastically change the point of view by treating a non-selective measurement occurrence of the system $S$ as a one-time Poisson process (see Fig.3).
Namely, we forbid the direct use of the non-selective measurement process (\ref{eq:step2}), as in the GRW model\cite{GRW}, and treat a non-selective measurement occurrence statistically in the enlarged statistical ensemble whose elements themselves are statistical ensembles, each of which can be described by the conventional density matrix $\widehat{\varrho}$ that has been used already.
We distinguish the density matrix that describes this enlarged statistical ensemble (denoted by $\widehat{\rho}$) from the conventional density matrix (denoted by $\widehat{\varrho}$).
Whereas the density matrix $\widehat{\varrho}$ {\it directly} expresses one non-selective measurement occurrence in the individual ensemble with the individual definite elapsed time $\Delta \tau=\delta \tau$ to occur, the density matrix $\widehat{\rho}$ {\it statistically} expresses one non-selective measurement occurrence in the enlarged ensemble with the arithmetic average $\Delta \tau^{({\rm ave})}=\delta \tau$ of the individual elapsed times to occur over this enlarged ensemble
\begin{eqnarray}
\Delta \tau&=&\delta \tau\;,\\
\Delta \tau^{(\rm ave)}&\equiv&\int_0^\infty \tau^\prime w^{(0)}(\tau^\prime)d\tau^\prime \label{eq:TauAve}\\
&=&\delta \tau\;,\\
w^{(0)}(\tau)&\equiv&\frac{1}{\delta\tau}e^{-\tau/\delta\tau}\;.
\end{eqnarray}
In this, $w^{(0)}(\tau)$ is the {\it exponentially decaying} normalized distribution of elapsed time to occur for a non-selective measurement as a one-time Poisson process with a characteristic time $\delta \tau$ (see Fig.4).
Schematically, in the descriptions by the density matrices $\widehat{\varrho}^X(\tau)$ and $\widehat{\rho}^X(\tau)$ of a system $X$ (in our process, $X=S+M$), whether one non-selective measurement occurs, at $\tau=\delta\tau$ for $\widehat{\varrho}^X(\tau)$ and within $0<\tau\le \delta\tau$ for $\widehat{\rho}^X(\tau)$, in the statistical ensembles of the sample systems with pure states (i.e., whether the events become mutually exclusive, at $\tau=\delta\tau$ for $\widehat{\varrho}^X(\tau)$ and within $0<\tau\le \delta\tau$ for $\widehat{\rho}^X(\tau)$) can be expressed as
\begin{eqnarray}
\widehat{\varrho}^X(\tau=\delta\tau)&\Leftrightarrow&\overbrace{[X_1,X_2,\ldots,X_N]}^{\rm Yes}\;,\label{eq:XX0}\\
\widehat{\rho}^X(0<\tau\le \delta\tau)&\Leftrightarrow&\Bigl[\overbrace{\Bigl[X^{(1)}_1,X^{(1)}_2,\ldots,X^{(1)}_N\Bigr]}^{\rm Yes\ or\ No},\overbrace{\Bigl[X^{(2)}_1,X^{(2)}_2,\ldots,X^{(2)}_N\Bigr]}^{\rm Yes\ or\ No},\ldots\nonumber\\
&&\ldots,\overbrace{\Bigl[X^{(M)}_1,X^{(M)}_2,\ldots,X^{(M)}_N\Bigr]}^{\rm Yes\ or\ No}\Bigr]\label{eq:XX1}\;,
\end{eqnarray}
where Yes-ensemble is an exclusive mixture and No-ensemble is a general mixture
\begin{eqnarray}
\overbrace{[X_1,X_2,\ldots,X_N]}^{\rm Yes}&
{=}&[|x_1\ra,|x_2\ra,\ldots,|x_N\ra]\;,\\
\overbrace{[X_1,X_2,\ldots,X_N]}^{\rm No}&
{=}&[|\Psi_1\ra,|\Psi_2\ra,\ldots,|\Psi_N\ra]\;.\label{eq:XXX}
\end{eqnarray}
Here, we introduce sample systems with pure states $X_i$ and $X_i^{(j)}$ ($i=1,2,\ldots,N$; $j=1,2,\ldots,M$), natural numbers $M,N\gg1$, $\widehat{{\mathscr{O}}}_X$-eigenstate vectors $|x\ra$ and normalized state vectors $|\Psi\ra$.
We take limits $M,N\to \infty$.

Now, we let $\delta \tau$ be the infinitesimal $d\tau_0$ (i.e., we let $\Delta \tau^{({\rm ave})}$ in Eq.(\ref{eq:TauAve}) be $d\tau_0$) to make the one-time Poisson process be an instantaneous inhomogeneous process.
  This operation implies three facts about the statistical description of one non-selective measurement occurrence by using the density matrix $\widehat{\rho}^X$.
First, the non-selective measurement {\it must} occur for all ensemble-elements within an infinitesimal time interval starting from $\tau=0$ (namely, before a finite time elapses since $\tau=0$).
Second, the actual occurrence time {\it for each individual ensemble-element} is randomly chosen within this time interval by a statistical law of the exponential population decay (see Fig.4).
Third, the occurrence time {\it averaged} over all ensemble-elements is a definite time $\tau=d\tau_0$.
Since the non-selective measurement for each individual ensemble-element {\it occurs only once}, its occurrence time is an {\it event} in the probabilistic sense.
So, we have cut off the one-time Poisson process of one non-selective measurement at this averaged occurrence time: this averaging operation is the meaning of the `{\it statistical description}' of one non-selective measurement occurrence.
In the resultant process, the off-diagonal part of the density matrix $\widehat{\rho}_{\rm od}^X(\tau)$ evolves with respect to one non-selective measurement in the same way as for radioactive decay as
\begin{equation}
-\frac{\pa \widehat{\rho}_{\rm od}^X(\tau)}{\pa \tau}=\delta(\tau)\widehat{\rho}_{\rm od}^X(\tau)\;.
\end{equation}

Here, we assume the next quantum mechanical equivalence, which is compatible with the event reading process.
\begin{itemize}
\item[A1]
The pair of the Hilbert space ${\cal V}^X$ of the state vectors and the space of the observables $\{\widehat{{\cal O}}_X\}$ of the system $X$ for $\widehat{\rho}^X(\tau)$ is always the same as that for $\widehat{\varrho}^X(\tau)$, up to the unitary equivalence
\begin{equation}
({\cal V}^X,\{\widehat{{\cal O}}_X\})\overset{\widehat{U}}{\simeq}(\widehat{U}{\cal V}^X,\{\widehat{U}\widehat{{\cal O}}_X\widehat{U}^{-1}\})\label{eq:XEq}
\end{equation}
for a unitary operator $\widehat{U}$.
\end{itemize}
For the basis of this assumption A1, see the explanations in Appendix A.

Next, the time evolution of the density matrix $\widehat{\rho}(\tau)$ of the total system $S+M$ before the event reading (i.e., $0\le\tau<\tau(t_0)$) consists of two parts.
Before and after step (ii), the time evolution of the density matrix $\widehat{\rho}(\tau)$ follows the conventional von Neumann equation (that is, the Schr${\ddot{{\rm o}}}$dinger equation for its matrix elements\cite{GRW}).
As a result of the non-selective measurement (i.e., step (ii)), with respect to the double kets, the off-diagonal part of the density matrix $\widehat{\rho}_{\rm od}(\tau)$ changes by a multiplicative factor $e^{-1}$ in the same way as for radioactive decay, while the diagonal part does not change.
In the following, we see this fact by solving the reduced von Neumann equation during $0\le\tau<\tau(t_0)$ (the solution is Eq.(\ref{eq:int0})).

The reduced von Neumann equation for the cut-off inhomogeneous one-time Poisson process of one instantaneous non-selective measurement takes the form
\begin{eqnarray}
\widehat{\rho}(\tau+d\tau)&=&(1-\delta(\tau)d\tau)\biggl(\widehat{\rho}(\tau)-\frac{i}{\hbar}[\widehat{{\cal H}}_{\rm kin}+\widehat{{\cal H}}_{\rm int}(\tau),\widehat{\rho}(\tau)]d\tau\biggr)\nonumber\\
&&
+\delta(\tau)d\tau\sum_{{\rm all}\ y}\widehat{P}(y)\widehat{\rho}(\tau)\widehat{P}(y)\;.\label{eq:world}
\end{eqnarray}
Here, the two factors, $(1-\delta(\tau)d\tau)$ and $\delta(\tau)d\tau$ are treated as $1$ and $0$, respectively, when $\tau\neq 0$; and are treated as $\epsilon_0$ (a positive infinitesimal) and $1-\epsilon_0$, respectively, when $\tau=0$.
In this section, starting from this equation, $\widehat{{\cal H}}_{\rm kin}$ and $\widehat{{\cal H}}_{\rm int}(\tau)$ refer to the Hamiltonians after the contraction of the time interval $I$ to an instant $\tau=d\tau_0$ on average and their time range is divided into $0\le \tau\le d\tau_0$ and $d\tau_0<\tau<\tau(t_0)$.
In particular,
\begin{equation}
\widehat{{\cal H}}_{\rm int}(\tau)=\left\{\begin{array}{cc}0&0\le \tau\le d\tau_0\\
&\\
\widehat{{\cal H}}_{\rm fb}(\widehat{{\mathscr{O}}}^{S_0}\otimes \widehat{1}^A\otimes \widehat{1}^M)&d\tau_0<\tau<\tau(t_0)\;.
\end{array}\right.\label{eq:HINT}
\end{equation}

We rewrite Eq.(\ref{eq:world}) as the differential equation
\begin{eqnarray}
\frac{\pa}{\pa \tau}\widehat{\rho}(\tau)
&=&
-\frac{i}{\hbar}[\widehat{{\cal H}}_{\rm kin}+\widehat{{\cal H}}_{\rm int}(\tau),\widehat{\rho}(\tau)]
\nonumber\\&&
-\delta(\tau)\Biggl(\widehat{\rho}(\tau)-\sum_{{\rm all}\ y}\widehat{P}(y)\widehat{\rho}(\tau)\widehat{P}(y)
\Biggr)\;.\label{eq:gen}
\end{eqnarray}
This equation is for the partial density matrix elements in the representation using two $\widehat{{\mathscr{O}}}$-coordinates $(x,y)$
\begin{eqnarray}
\frac{\pa}{\pa \tau}\La x|\widehat{\rho}(\tau)|y\Ra
&=&-\frac{i}{\hbar}\La x|[\widehat{{\cal H}}_{\rm kin}+\widehat{{\cal H}}_{\rm int}(\tau),\widehat{\rho}(\tau)]|y\Ra 
\nonumber\\
&&
-\delta (\tau)(1-\Delta_r(x-y))
\La x|\widehat{\rho}(\tau)|y\Ra \;,
\label{eq:LiouvilleEq0}\\
{{\Delta_r}}(x-y)&\equiv& \left\{\begin{array}{cc}1&x=y \\
& \\
0&x\neq y\;.
\end{array}\right.\label{eq:int}
\end{eqnarray}

By solving this equation, for $0\le \tau<\tau(t_0)$, we obtain
\begin{eqnarray}
\La x|\widehat{\rho}(\tau)|y\Ra= e^{-\theta(\tau)(1-{\Delta_r}(x-y))}\La x|\widehat{\rho}_{{\rm{Sch}}}(\tau)|y\Ra\label{eq:int0}
\end{eqnarray}
with $\theta(\tau)$ as the Heaviside unit step function that satisfies $d{\theta}(\tau)/d\tau=\delta(\tau)$.
[Note that $\Lambda^A$ and $\Lambda^M$ in $\widehat{{\cal H}}_{\rm int}(t)$ are strong enough that we can neglect $\widehat{{\cal H}}_{\rm kin}$ during steps (ii) and (iii) by using time parameters rescaled by the factors $\Lambda^A$ (for step (ii)) and $\Lambda^M$ (for step (iii)).]
Namely, by the non-selective measurement, only off-diagonal matrix elements of the density matrix $\widehat{\rho}(\tau)$ with respect to the double kets change.
So, the trace of the density matrix $\widehat{\rho}(\tau)$ over the total system remains unity.
This fact comes from the property of projection operator shown in Eq.(\ref{eq:P}).
In Eq.(\ref{eq:int0}), we introduced the density matrix $\widehat{\rho}_{{\rm{Sch}}}(\tau)$ of the total system ${{S}}+M$ in the {\it absence} of both a non-selective measurement and its subsequent event reading.
This density matrix satisfies the same initial conditions as $\widehat{\rho}(\tau)$ and the von Neumann equation for the density matrix (that is, the Schr$\ddot{{\rm{o}}}$dinger equation for its matrix elements\cite{GRW})
\begin{equation}
\frac{\pa}{\pa \tau}\widehat{\rho}_{{\rm{Sch}}}(\tau)=-\frac{i}{\hbar}[\widehat{{\cal H}}_{\rm kin}+\widehat{{\cal H}}_{\rm int}(\tau),\widehat{\rho}_{{\rm{Sch}}}(\tau)]\label{eq:SCHEQ}
\end{equation}
after the contraction of the time interval $I$ to an instant $\tau=d\tau_0$ on average.

\begin{figure}[htbp]
\begin{center}
\includegraphics[width=0.5\hsize,bb=0 0 260 290]{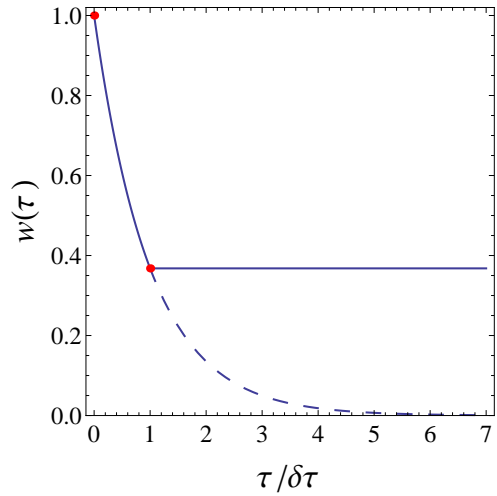}
\end{center}
\caption{
In Eq.(\ref{eq:world}), the characteristic time of the inhomogeneous one-time Poisson process is an infinitesimal $\delta\tau$.
The time-dependent fraction $w(\tau)$ of the ensemble-elements in the enlarged statistical ensemble described by the density matrix $\widehat{\rho}(\tau)$, to which {\it no} non-selective measurement of the system $S$ is applied until the time becomes $\tau$, is shown by the blue solid curve.
Here, $w(\tau)$ is the factor $e^{-\theta(\tau)(1-\Delta_r(x-y))}$ ($x\neq y$) in Eq.(\ref{eq:int0}).
The two red points are at $(0,1)$ and $(1,e^{-1})$.
The blue dashed curve indicates a part of the function $e^{-\tau/\delta\tau}$ of $\tau$.
For a finite elapsed time $\tau$, $e^{-\tau/\delta\tau}$ is zero.}
\end{figure}

Up to now, we have studied the time evolution of measurement in the total system ${{S}}+M$ in terms of its density matrix $\widehat{\rho}(\tau)$.
In the next subsection, we discuss the time evolution of the combined measured system ${{S}}$ and its external measuring system $M$ separately, motivating an axiomatic attachment of the factors $e^{-1}$ and $e^{\Delta_r}$ appearing in the right-hand side of Eq.(\ref{eq:int0}) at $\tau=d\tau_0$ to the subsystems ${{S}}$ and $M$.

To prepare for the following parts, here we introduce the superoperator $\widetilde{\Delta_r}$ acting on an arbitrary operator $\widehat{{\cal O}}_{{S}}$ of ${{S}}$ by
\begin{equation}
\widetilde{\Delta_r}\widehat{{\cal O}}_{{S}}\overset{\widetilde{\Delta_r}}{\equiv}\sum_{{\rm all}\ y}\widehat{P}(y)\widehat{{\cal O}}_{{S}}\widehat{P}(y)\;.
\end{equation}
This superoperator corresponds to $\Delta_r$ and acts on the density matrix and observables of ${{S}}$ (i.e., the non-selective measurement superoperator in the von Neumann equation, Eq.(\ref{eq:world})).

\subsection{Entropy transfer ${\sigma}_{\bar{Y}\to Y}$}

In this subsection, we assume $d\tau_0\le \tau <\tau(t_0)$.
By non-selective measurement of ${{S}}$, two factors arise for normalization, $e^{\Delta_r}$ and $e^{-1}$, in Eq.(\ref{eq:int0}) from the second term and the first term in the von Neumann equation (\ref{eq:world}), respectively.
The factor $e^{\Delta_r}$ definitely acts on the density matrix $\widehat{\rho}^S$ of ${{S}}$, but the factor $e^{-1}$ does {\it not}.

Regarding the problem of the attachment of the factor $e^{-1}$ to a subsystem of the system $S+M$, we first provide general and formal arguments and then we give an intuitive explanation for them.

First, incorporating these normalization factors $e^{\Delta_r}$ and $e^{-1}$, we consider the density matrix $\widehat{\rho}^Y$ of a subsystem $Y={{S}},M$ of the system ${{S}}+M$, applied to {\it its redefined} observables of $Y$ (see Eq.(\ref{eq:Key})) as a statistical operator, after an entropy transfer when we assume the event reading process subsequent to a non-selective measurement.
Then, we define the {\it entropy transfer}, ${\sigma}_{\bar{Y}\to Y}$, from the complementary system $\bar{Y}$ of the subsystem $Y$ in the total system $S+M$ to the system $Y$, accompanying the selective measurement, by the next relation:
\begin{eqnarray}
e^{{\sigma}_{\bar{Y}\to Y}}\widehat{\rho}^Y&\overset{{\sigma}_{\bar{Y}\to Y}}{\equiv}&\widehat{\rho}_{{0}}^Y\label{eq:Key0}\\
&\overset{\widehat{\rho}_{{0}}^Y}{\equiv}&{\rm tr}_{{\bar Y}}e^{-(1-\widetilde{\Delta_r})}\widehat{\rho}_{{\rm Sch}}\;.\label{eq:Key00}
\end{eqnarray}
Here, the condition (refer to Eq.(\ref{eq:Proof}))
\begin{equation}
{\sigma}_{\bar{Y}\to Y}=-{\sigma}_{Y\to \bar{Y}}\label{eq:RR}
\end{equation}
is satisfied.
From these definitions, we note that
\begin{eqnarray}
{\rm tr}_Y\widehat{{\rho}}_{{0}}^Y&=&1\;,\label{eq:RhoSch}\\
{\rm tr}_Y\widehat{\rho}^Y&=&e^{-{\sigma}_{\bar{Y}\to Y}}\;.\label{eq:Rho}
\end{eqnarray}
It follows from Eq.(\ref{eq:Rho}) that, for a finite ${\sigma}_{\bar{Y}\to Y}$, ${\rm tr}_Y[\widehat{1}_Y\widehat{\rho}^Y]\neq 1$ holds.
(However, for $\widehat{1}_Y^\star$, introduced later, ${\rm tr}_Y[\widehat{1}_Y^\star \widehat{\rho}^Y]=1$ holds.)
That is, for such a case, $\widehat{\rho}^Y$ is ill-defined (of course, $\widehat{\rho}_{0}^Y$ is well-defined from Eq.(\ref{eq:RhoSch})) for use as the statistical operator for the observables $\{\widehat{{\cal O}}_Y\}$ defined in the absence of the event reading.
For an arbitrary observable $\widehat{{\cal O}}_Y$ of $Y$ defined in the absence of the event reading, we can define a corresponding operator $\widehat{{\cal O}}_Y^{\star}$ which differs from $\widehat{{\cal O}}_Y$ by at most multiplication by a $c$-number and to which $\widehat{\rho}^Y$ is applied as the statistical operator.
This operator $\widehat{{\cal O}}_Y^{\star}$ is defined by
\begin{eqnarray}
{\rm tr}_Y[\widehat{{\cal O}}_Y^\star\widehat{\rho}^Y]&\overset{\widehat{{\cal O}}_Y^\star}{\equiv}&\la{\widehat{{\cal O}}_Y}\ra\label{eq:Key1}\\
&\overset{\la{\widehat{{\cal O}}_Y}\ra}{\equiv}&{\rm tr}_Y[\widehat{{\cal O}}_Y\widehat{\rho}_{{0}}^Y]\;.\label{eq:Key2}
\end{eqnarray}
The correspondence between $\widehat{{\cal O}}_Y$ and $\widehat{{\cal O}}_Y^\star$ is well-defined because the quantum statistical average $\la{\widehat{{\cal O}}_Y}\ra$ is well-defined for both of the pairs $(\widehat{\rho}^Y_{0},\widehat{{\cal O}}_Y)$ and $(\widehat{\rho}^Y,\widehat{{\cal O}}_Y^\star)$.
From Eqs.(\ref{eq:Key0}), (\ref{eq:Key1}) and (\ref{eq:Key2}), the transformation rule for the observable $\widehat{{\cal O}}_Y$ by the entropy transfer
\begin{equation}
\widehat{{\cal O}}_Y^\star=e^{{\sigma}_{\bar{Y}\to Y}}\widehat{{\cal O}}_Y \label{eq:Key}
\end{equation}
follows.
By this relation, we find that the condition (\ref{eq:RR}) is equivalent to
\begin{equation}
{\rm tr}_{Y+\bar{Y}}\Bigl[\Bigl(\widehat{{1}}_Y^\star\otimes \widehat{{1}}_{\bar{Y}}^\star\Bigr)\widehat{\rho}\Bigr]=1\;.\label{eq:Proof}
\end{equation}
We will use the relation given by Eq.(\ref{eq:Key}) in Sec. 3.

Now, we give an intuitive explanation for the formal arguments in Eqs.(\ref{eq:RhoSch}) to (\ref{eq:Key}).
For the set of observables $\{\widehat{{\cal O}}_Y\}$, the density matrix $\widehat{\rho}^Y_{0}$ can be interpreted as a mixed ensemble, ${\mathfrak E}_Y(\{(|\Psi\ra,w)\})$, of a large number of samples of the system $Y$, described in terms of subensembles of normalized state vectors $\{|\Psi\ra\}$ with statistical probabilities (i.e., fractions) $\{w\}$ such that $\sum_nw_n=1$:
\begin{equation}
\widehat{\rho}^Y_{0}=\sum_nw_n|\Psi_n\ra\la \Psi_n|\;,\ \{\widehat{{\cal O}}_Y\}\;.\label{eq:RHO0}
\end{equation}
The basis for this interpretation of $\widehat{\rho}^Y_{0}$ is that, for an arbitrarily given observable $\widehat{{\cal O}}_Y$ of the system $Y$ having eigenvalues $\{{\cal O}_{Y}\}$ and corresponding eigenvectors $\{|{\cal O}_{Y}\ra\}$, this density matrix $\widehat{\rho}^Y_{0}$ gives the ensemble average of the measurement results of $\widehat{{\cal O}}_Y$ measured over all sample systems in the statistical ensemble ${\mathfrak E}_Y(\{(|\Psi\ra,w)\})$ by the trace operation:
\begin{eqnarray}
\sum_{s,n}w_n|\la {\cal O}_{Y,s}|\Psi_n\ra|^2{\cal O}_{Y,s}={\rm tr}_Y[\widehat{{\cal O}}_Y\widehat{\rho}^Y_{0}]\;.\label{eq:OY}
\end{eqnarray}
Here, the Born rule in the quantum measurement is applied to Eq.(\ref{eq:OY}) as the quantum mechanical probability factors $|\la {\cal O}_{Y}|\Psi\ra|^2$ (i.e., the occurrence rates of the measurement outcomes of $\widehat{{\cal O}}_Y$ for subensembles of pure states $|\Psi\ra$) in Eq.(\ref{eq:OY}).
Now, for the same reason, the density matrix $\widehat{\rho}^Y$ can be interpreted as the same mixed ensemble ${\mathfrak E}_Y(\{(|\Psi^\star\ra,w)\})$ of systems for the sets of normalized state vectors $\{|\Psi^\star\ra\}$ equivalent to $\{|\Psi\ra\}$ and redefined observables $\{\widehat{{\cal O}}_Y^\star\}$ defined by Eq.(\ref{eq:Key}):
\begin{eqnarray}
\widehat{\rho}^Y&=&\sum_nw_ne^{-\sigma_{\bar{Y}\to Y}}|\Psi_n\ra\la \Psi_n|\;,\ \{\widehat{{\cal O}}_Y^\star\}\\
&=&\sum_nw_n|\Psi_n^\star\ra\la \Psi_n^\star|\;,\ \{\widehat{{\cal O}}_Y^\star\}\;.\label{eq:RHO}
\end{eqnarray}
This is because $\widehat{\rho}^Y$ gives the correct occurrence rates of the measurement outcomes of $\widehat{{\cal O}}_Y^\star$ measured over all sample systems in the statistical ensemble ${\mathfrak E}_Y(\{(|\Psi^\star\ra,w)\})$ by the trace operation of Eq.(\ref{eq:Key1}).

For the discussion in Sec. 3, here it must be noted that the entropy transfer $\sigma_{\bar{Y}\to Y}$ is also valid in the description when using the density matrix $\widehat{\varrho}^Y$.
This is because the change of the pair of the Hilbert space ${\cal V}^Y$ of the state vectors and the space of the observables $\{\widehat{{\cal O}}_Y\}$ of the system $Y$ from $({\cal V}^Y,\{\widehat{{\cal O}}_Y\})$ to $({\cal V}^{Y\star},\{\widehat{{\cal O}}_Y^\star\})$ is common between $\widehat{\rho}^Y$ and $\widehat{\varrho}^Y$, up to the unitary equivalence 
\begin{equation}
({\cal V}^Y,\{\widehat{{\cal O}}_Y\})\overset{\widehat{U}}{\simeq}(\widehat{U}{\cal V}^Y,\{\widehat{U}\widehat{{\cal O}}_Y\widehat{U}^{-1}\})\label{eq:YEq}
\end{equation}
(see assumption A1 in Sec.2.2).

In Sec. 2.4, we will determine the values of ${\sigma}_{\bar{Y}\to Y}$ in Eqs.(\ref{eq:Key0}) and (\ref{eq:Key}) (see Eqs.(\ref{eq:ResultR}) and (\ref{eq:ResultR2}) for the results in the case of the main statement).

The entropy production in the main statement of this paper is made on the combined measured system ${{S}}$ by a measuring system $M$ that is external to the system ${{S}}$.
We specifically consider the case of $Y={{S}}$ in the above definitions.
From Eq.(\ref{eq:int0}) (or Eq.(\ref{eq:RR})), in the total system ${{S}}+M$, there is no {\it net} entropy production in the measurement process.
However, in the total system ${{S}}+M$, when we do a measurement including the selection of a quantum eigenstate primarily of the measuring system from an exclusive mixture, there arises an internal transfer of entropy (in another term, {\it pair entropy production}) required by the measuring system $M$ to the combined measured system ${{S}}$ inside the total system ${{S}}+M$.
This means, as its role will be revealed in the next section, that for the entropy transfer ${\sigma}_{M\to {{S}}}$ and the entropy production ${\sigma}$ accompanying the measurement process,
\begin{eqnarray}
{\sigma}_{M\to {{S}}}&\neq& 0\;,\\
{\sigma}_{{{S}}+M}&=&{\sigma}_{M\to {{S}}}+{\sigma}_{{{S}}\to M}\\
&=&0
\end{eqnarray}
hold.

Here, for clarity, we note that the transferred entropy $\sigma_{\bar{Y}\to Y}$ explained up to this point is independent of the thermodynamic entropy, $H(\{p_n\})$, occurring under Landauer's principle for information erasure\cite{Landauer}.
($H(\{p_n\})$ denotes the Shannon entropy $-\sum_np_n\ln p_n$ of a given full set of probabilities $\{p_n\}$ of measurement outcomes $\{n\}$.)
In the following explanations, we omit the heat bath.
In the quantum thermodynamics of information\cite{SU2}, it was shown that the thermodynamic entropy $H(\{p_n\})$ occurring under Landauer's principle for information erasure is attributed to a mathematical identity involving the von Neumann entropy $S(\widehat{\varrho}^{\cal M})\equiv-{\rm tr}(\widehat{\varrho}^{\cal M}\ln \widehat{\varrho}^{\cal M})$ of the memory system ${\cal M}$ that stores information on the outcome of a measurement:
\begin{equation}
H(\{p_n\})=S\bigl(\widehat{\varrho}_0^{\prime{\cal M}}\bigr)-\sum_np_nS\bigl(\widehat{\varrho}_n^{{\cal M}}\bigr)\;.\label{eq:Landau}
\end{equation}
This mathematical identity holds under the direct sum structure of the memory state space ${\cal V}^{\cal M}$ with respect to the label $n$ of the memory states (here, $n=0$ represents the standard memory state):
\begin{equation}
{\cal V}^{\cal M}=\bigoplus_n {\cal V}^{\cal M}_n\;.\label{eq:Str}
\end{equation}
In Eq.(\ref{eq:Landau}), the supports of $\widehat{\varrho}_n^{{\cal M}}$ and $\widehat{\varrho}_0^{\prime{\cal M}}$ belong to ${\cal V}^{\cal M}_n$ and ${\cal V}_0^{\cal M}$, respectively, with unit probability, and $\widehat{\varrho}_0^{\prime{\cal M}}$ is a unitary transformation of the state $\sum_np_n\widehat{\varrho}_n^{\cal M}$.
Indeed, when $\widehat{\varrho}_n^{\cal M}$ are the canonical distributions $\widehat{\varrho}_{n,{\rm can}}^{\cal M}$, due to Klein's inequality $-{\rm tr}(\widehat{\varrho}_0^{\prime{\cal M}}\ln \widehat{\varrho}^{\cal M}_{0,{\rm can}})\ge S(\widehat{\varrho}_0^{\prime{\cal M}})$, Eq.(\ref{eq:Landau}) times $k_BT$ is not greater than sum of the work $W_{\rm eras}$ and the free energy difference $\Delta F$ that accompany the information erasure process
\begin{equation}
\sum_np_n\widehat{\varrho}_{n,{\rm can}}^{\cal M}\longrightarrow\widehat{\varrho}_0^{\prime{\cal M}}\;.\label{eq:ier}
\end{equation}
From this result, when the free energy difference $\Delta F$ is zero (for instance, in the case of a symmetric potential memory), Landauer's principle $W_{\rm eras}\ge k_BTH(\{p_n\})$ follows\cite{SU2}.
In summary, $H(\{p_n\})$ comes from a state change (\ref{eq:ier}) under the structure to store information, that is, Eq.(\ref{eq:Str}).
In contrast, our transferred entropy $\sigma_{\bar{Y}\to Y}$ is independent of the structure used to store information (i.e., Eq.(\ref{eq:Str})) and comes from the finite variation of the logarithm of the normalization constant of the density matrix (see footnote $\S$).
So, the two entropies $\sigma_{\bar{Y}\to Y}$ and $H(\{p_n\})$ have mutually independent physical origins.
In particular, $\sigma_{\bar{Y}\to Y}$ does not have an informatical origin; instead, it is attributed to the population decay (see Fig.4) in the enlarged statistical ensemble (\ref{eq:XX1}) described by $\widehat{\rho}^X$ ($X=S+M$) due to the reduction of our knowledge about the system $X$.

\subsection{Definition of a measuring system $M$}

In this subsection, we precisely define a measuring system $M$ that can read the event after a non-selective measurement.

We define the measured system $S$ as the system in which the non-selective measurement part (i.e., the superoperator $\widetilde{\Delta_r}$) acts, and {\it if $S$ is an isolated system, then $S$ cannot complete the selective measurement}.

Before we can state the definition of $M$, we need a few preliminaries.

In the following, we call two states {\it unitarily equivalent} to each other when these states are related by a unitary transformation $\widehat{U}$.
Here, the unitary transformation $\widehat{U}$ for the equivalence is treated as a wide-sense (ws) transformation: it transforms both the state vector $|\psi\ra$ and all of the observables $\{\widehat{{\cal O}}\}$, simultaneously, as 
\begin{equation}
{\rm ws}:|\psi\ra\to \widehat{U}|\psi\ra\;,\ \ \widehat{{\cal O}}\to \widehat{U}\widehat{{\cal O}}\widehat{U}^{-1}
\end{equation}
in the sense of Dirac's transformation theory.
Since a unitary transformation of this kind does not change the spectra of all observables, the superposition relation and inner product between states or, in general, arbitrary algebraic relations between observables and state vectors, it changes no content of the quantum mechanics.
So, {\it elements belonging to the same wide-sense unitary equivalence class have the same quantum mechanical contents.}
For contrast, we introduce the unitary equivalence in a narrow sense (ns) via the unitary transformation
\begin{equation}
{\rm ns}:|\psi\ra\to \widehat{U}_1|\psi\ra\;,\ \ \widehat{{\cal O}}\to\widehat{U}_2\widehat{{\cal O}}\widehat{U}_2^{-1}
\end{equation}
for unitary transformations $\widehat{U}_1$ and $\widehat{U}_2$.
Unitarily equivalent states (in both senses) form an equivalence class because the product of two unitary transformations is unitary and the inverse of a unitary transformation is unitary.

We now introduce an assumption about the event reading process.
\begin{itemize}
\item[A2] Two states that are unitarily equivalent in the wide sense at $\tau=\tau(t_0)-\epsilon$ are unitarily equivalent in the wide sense at $\tau=\tau(t_0)$.
This equivalence at $\tau=\tau(t_0)$ also implies stochasticity obeying the Born rule if event reading occurs at $\tau=\tau(t_0)$.
\end{itemize}
That is, the event reading process is compatible with quantum mechanics formalism, and its cause is the occurrence time, rather than the state.

We consider the state of the system $S+M$ at $\tau=\tau(t_0)-\epsilon$ (i.e., at the time when the non-selective measurement of the system $S+M$ completes):
\begin{equation}
\widehat{\rho}(\tau=\tau(t_0)-\epsilon)\;,\ \{\widehat{{\cal O}}^\star_{S+M}\}\;.\label{eq:No1}
\end{equation}

Since time evolution $\widehat{U}_{{\rm fb}}$ driven by the entangling interaction is unitary, by reversing this process as $\widehat{U}^{-1}_{\rm fb}$, we find that Eq.(\ref{eq:No1}) is unitarily equivalent to
\begin{equation}
\widehat{\rho}(\tau=d\tau_0)=\widehat{U}_{\rm fb}^{-1}\widehat{\rho}(\tau=\tau(t_0)-\epsilon)\widehat{U}_{\rm fb}\;,\ \{\widehat{U}^{-1}_{\rm fb}\widehat{{\cal O}}^\star_{S+M}\widehat{U}_{\rm fb}\}\label{eq:No2}
\end{equation}
in the wide sense.
In Eq.(\ref{eq:No2}), the systems $S$ and $M$ are decoupled from each other; specifically, in the statistical description of the states by the statistical ensembles, the systems $S$ and $M$ have no statistical correlation.
Thus, we can separate the systems $S$ and $M$ from each other.
The states of the systems $S$ and $M$ after separation are
\begin{eqnarray}
\widehat{\rho}^S(\tau=d\tau_0)&=&{\rm tr}_M^\star \widehat{\rho}(\tau=d\tau_0)\;,\ \{{\rm tr}_M^\star[\widehat{U}^{-1}_{\rm fb}\widehat{{\cal O}}^\star_{S+M}\widehat{U}_{\rm fb}\widehat{\rho}^M]\}\;,\label{eq:No33a}\\
\widehat{\rho}^M(\tau=d\tau_0)&=&{\rm tr}_S^\star \widehat{\rho}(\tau=d\tau_0)\;,\ \{{\rm tr}_S^\star[\widehat{U}^{-1}_{\rm fb}\widehat{{\cal O}}^\star_{S+M}\widehat{U}_{\rm fb}\widehat{\rho}^S]\}\;,
\end{eqnarray}
(the partial trace ${\rm tr}_Y^\star$ is taken for the redefined state (\ref{eq:RHO})).
These density matrices in the case of no entropy transfer $\sigma_{M\to S}$ reduce to the {\it original partial traces} of the density matrix $\widehat{\rho}$ in Eq.(\ref{eq:No2}):
\begin{eqnarray}
\widehat{\rho}^S_0(\tau=d\tau_0)&=&{\rm tr}_Me^{-(1-\widetilde{\Delta_r})}\widehat{\rho}(\tau=0)\;,\ \{{\rm tr}_M[\widehat{U}^{-1}_{\rm fb}\widehat{{\cal O}}_{S+M}\widehat{U}_{\rm fb}\widehat{\rho}_0^M]\}\;,\label{eq:No3a}\\
\widehat{\rho}^M_0(\tau=d\tau_0)&=&{\rm tr}_Se^{-(1-\widetilde{\Delta_r})}\widehat{\rho}(\tau=0)\;,\ \{{\rm tr}_S[\widehat{U}^{-1}_{\rm fb}\widehat{{\cal O}}_{S+M}\widehat{U}_{\rm fb}\widehat{\rho}_0^S]\}\;.\label{eq:No3b}
\end{eqnarray}
Namely, the density matrix $\widehat{\rho}^Y_0$ ($Y=S,M$) matches that of the system $Y$ {\it when we remove the dynamical degrees of freedom of its complementary system $\bar{Y}$ from the total system $S+M$}.
Here, we regard the density matrix $\widehat{\rho}_0^Y$ ($Y=S,M$) as a statistical ensemble ${\mathfrak E}_Y(\{(|\Psi\ra,w)\})$, as in Eq.(\ref{eq:RHO0}).
In the definition of the density matrix $\widehat{\rho}^Y_0$ ($Y=S,M$), $M$ is an {\it ordinary system}, that is, a system interacting with $S$ unitarily and evolving unitarily by itself.

In Eqs.(\ref{eq:No33a}) to (\ref{eq:No3b}), we use the following argument.
The set of well-defined observables $\{\widehat{{\cal O}}^{({\rm w-d})}_Y\}$ of the system $Y$ ($Y=S,M$) changes after {the entropy transfer is made at time $\tau_{\rm et}$ satisfying $d\tau_0\le \tau_{\rm et}<\tau(t_0)$}:
\begin{eqnarray}
\{\widehat{{\cal O}}^{({\rm w-d})}_Y\}=\left\{\begin{array}{cc}
\{\widehat{{\cal O}}_Y\}&\tau< \tau_{\rm et}\\
& \\
\{\widehat{{\cal O}}_Y^\star\}&\tau\ge \tau_{\rm et}\;.\end{array}\right.
\end{eqnarray}
However, whenever the system $M$ is an {\it ordinary system},
\begin{equation}
\{\widehat{{\cal O}}_Y^{({\rm w-d})}\}\equiv \{\widehat{{\cal O}}_Y\}
\end{equation}
holds.

Now, the definition of $M$ consists of two conditions (refer to Figs.5 and 6).
\begin{figure}[htbp]
\begin{center}
\includegraphics[width=0.55\hsize,bb=0 0 260 170]{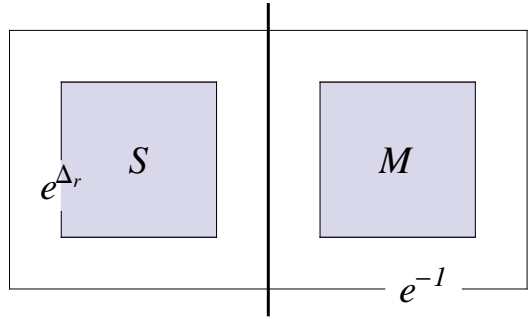}
\end{center}
\caption{
This figure schematically shows the factorization of the normalization factor $e^{-(1-\Delta_r)}$ in Eq.(\ref{eq:int0}) into the systems ${S}$ and $M$ at $\tau=\tau_{\rm et}$ such that $M$ {\it is a measuring system}.
In this factorization, while the entropy production on the system $S$ separated from the system $M$ is $-\Delta_r(0)=-1$, the entropy production on the system $S$ obtained by removing the dynamical degrees of freedom of the system $M$ from the total system $S+M$ is $1-\Delta_r(0)=0$.
This means that the removal of the dynamical degrees of freedom of the system $M$ from the total system $S+M$ changes the wide-sense unitary equivalence class of the state of the system $S$.
In this sense, {\it the system $S$ is not isolated}.
It is obvious that this property is true for the system $M$ also.}
\end{figure}

\begin{figure*}[htbp]
\includegraphics[width=0.55\hsize,bb=0 0 260 170]{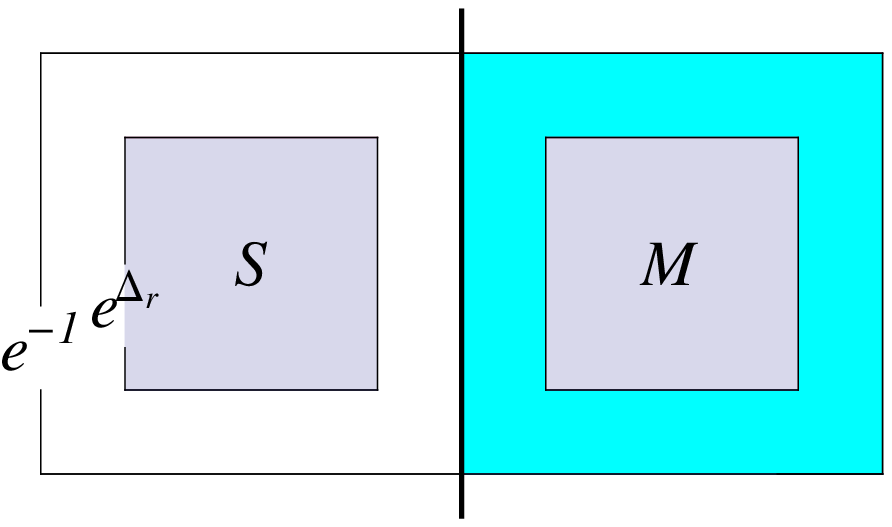}
\includegraphics[width=0.55\hsize,bb=0 0 267 177]{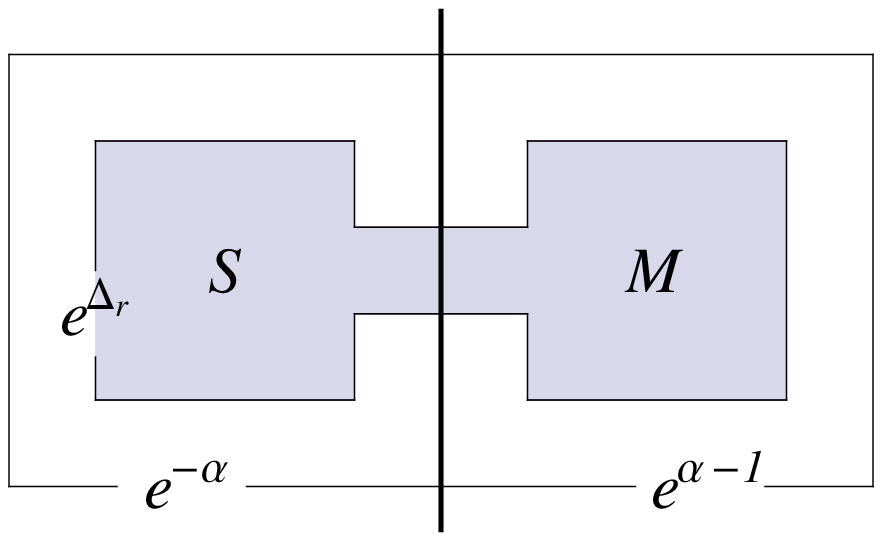}
\caption{
Both figures schematically show factorizations of the normalization factor $e^{-(1-\Delta_r)}$ in Eq.(\ref{eq:int0}) into the systems ${S}$ and $M$ such that $M$ {\it does not satisfy} the conditions on the measuring system.
In the left-hand figure, the process of non-selective measurement of ${{S}}$ as a preparation for the selective measurement is a closed process in $S$, and $S$ is isolated.
Then, $M$ cannot read the measurement result ${\mathfrak M}_x$: namely, $M$ cannot do a selective measurement (this result is shown in the main text).
In the right-hand figure with $\alpha \neq 0,1$, the condition of the independence of the systems ${{S}}$ and $M$ is not satisfied.}
\end{figure*}

We state the first condition. 
\begin{itemize}
\item[B1]
In the total system ${{S}}+M$, $M$ and ${{S}}$ are {\it independent} from each other as quantum systems in the sense that any time-dependent process in the total system ${{S}}+M$ is divisible into parts for ${{S}}$ and $M$.
\end{itemize}

The precise definition of the {\it independence} of two systems $S_1$ and $S_2$ consists of two conditions.
First, there is no overlapping degree of freedom between $S_1$ and $S_2$.
Second, any time-dependent process in the combined system $S_1+S_2$ can be decomposed as a combination of transitions between {\it elements} (defined below) and any transition appearing in this decomposition is either a transition between two elements of $S_1$ or a transition between two elements of $S_2$.
Here, an {\it element} refers to $\{x\}$ or $\{y\}$ which is a set of labels fully distinguishing diagonal elements of the density matrix $\widehat{\rho}^{S_1}$ of $S_1$ or the density matrix $\widehat{\rho}^{S_2}$ of $S_2$ under the time-dependent spectral decomposition
\begin{eqnarray}
\widehat{\rho}^{S_1}&=&\sum_{\{x\}}p_1(\{x\})|\Psi_1(\{x\})\ra\la \Psi_1(\{x\})|\;,\\
\widehat{\rho}^{S_2}&=&\sum_{\{y\}}p_2(\{y\})|\Psi_2(\{y\})\ra\la \Psi_2(\{y\})|\;.
\end{eqnarray}

We state the second condition.
\begin{itemize}
\item[B2] {\it If} at $\tau=\tau(t_0)-\epsilon$, we can separate the systems $S$ and $M$ from each other and the unitary equivalence class of the state of the system $S$ in the wide sense is the same as the (wide-sense) unitary equivalence class of the state (\ref{eq:No3a}) obtained by the original partial trace of Eq.(\ref{eq:No2}) with respect to the system $M$ (i.e., by removing the dynamical degrees of freedom of the system $M$ from the total system $S+M$), then the (narrow-sense) unitary equivalence class of the state of the system $S+M$ at $\tau=\tau(t_0)$ does not change from that of the state of the system $S+M$ at $\tau=\tau(t_0)-\epsilon$.
\end{itemize}
The ground for this condition is that the form of Eq.(\ref{eq:step3}) as an exclusive mixture of product states implies that we cannot, in principle, tell the system $S$ from the system $M$ as the cause of the event reading (\ref{eq:step4}) that occurred.
If the assumption of this condition holds (then, the systems $S$ and $M$ are isolated from each other at $\tau=\tau(t_0)-\epsilon$\footnote{In connection with the definition of an {\it isolated state}, we assume that causality in the total system $S+M$ is reset just after an event reading as an acausal change.
After this, the systems $S$ and $M$ are in their isolated states again and the rule of causal, continuous and reversible change is applied to the systems $S$ and $M$.}\footnote{We denote by $\widehat{U}_S$ and $\widehat{U}_M$ the unitary operators of time evolution of the systems $S$ and $M$, respectively, driven by the kinetic Hamiltonians $\widehat{{\cal H}}^S_{\rm kin}$ and $\widehat{{\cal H}}^M_{\rm kin}$, again respectively.
If there is neither an event reading nor an entropy transfer, then the system $M$ described by using the density matrix, $\widehat{\varrho}^M(\tau=d\tau_0)$, in the same way as in Eq.(\ref{eq:No3b}) unitarily evolves from $\tau=\tau(t_0)-\epsilon$ as an isolated system.
(Note that, in this description, the systems $S$ and $M$ are decoupled but not statistically independent from each other.)
This is because
${\rm tr}_{S+M}[\widehat{U}_{\rm fb}^{-1}(\widehat{U}_S^{-1}\otimes \widehat{U}_M^{-1})(\widehat{1}_S\otimes \widehat{{\cal O}}_M)(\widehat{U}_S\otimes \widehat{U}_M)\widehat{U}_{\rm fb}\widehat{\varrho}(\tau=d\tau_0)]
={\rm tr}_{S+M}[(\widehat{1}_S\otimes \widehat{U}_M^{-1})(\widehat{1}_S\otimes\widehat{{\cal O}}_M)(\widehat{1}_S\otimes\widehat{U}_M)\widehat{U}_{\rm fb}(\sum_n\alpha_n|x_n,{\mathfrak A}_0\ra\la x_n,{\mathfrak A}_0|\otimes \widehat{\varrho}^M(\tau=d\tau_0))\widehat{U}_{\rm fb}^{-1}]
={\rm tr}_M[\sum_n\alpha_n\widehat{U}_{{\rm fb},M}^{(n)-1}\widehat{U}_M^{-1}\widehat{{\cal O}}_M\widehat{U}_M\widehat{U}_{{\rm fb},M}^{(n)}\widehat{\varrho}^M(\tau=d\tau_0)]={\rm tr}_M[\widehat{U}_M^{-1}(\sum_n\alpha_n \widehat{U}_{{\rm fb},M}^{(n)-1}\widehat{{\cal O}}_M\widehat{U}_{{\rm fb},M}^{(n)})\widehat{U}_M\widehat{\varrho}^M(\tau=d\tau_0)]$ holds for an arbitrary observable $\widehat{{\cal O}}_M$ of the system $M$ (here, unitary operators $\widehat{U}_{{\rm fb},M}^{(n)}$ are defined in the second equality; if ${\mathfrak M}$ refers to a pointer position, the equality $[\widehat{U}_M,\widehat{U}_{{\rm fb},M}^{(n)}]=0$ due to $[\widehat{{\cal H}}^M_{\rm kin},\widehat{{\mathfrak P}}^M_c]=0$ is used in the third equality; if ${\mathfrak M}$ refers to energy, the equality $\widehat{U}_M\widehat{U}_{{\rm fb},M}^{(n)}\widehat{\varrho}^M(\tau=d\tau_0)\widehat{U}_{{\rm fb},M}^{(n)-1}\widehat{U}_M^{-1}=\widehat{U}_{{\rm fb},M}^{(n)}\widehat{U}_M\widehat{\varrho}^M(\tau=d\tau_0)\widehat{U}_M^{-1}\widehat{U}_{{\rm fb},M}^{(n)-1}$ is used in the third equality): this is the unitary time evolution of an isolated system in the Heisenberg picture.
It is obvious that this property is true for the system $S$ also.}) and the system $M$ reads an event at $\tau=\tau(t_0)$, then {\it the isolated system $S$ in fact reads the event from the exclusive mixture of the system $S$} at $\tau=\tau(t_0)$.
This consequent contradicts the definition of the system $S$.
Thus, if the assumption of this condition holds, the state of the system $S+M$ at $\tau=\tau(t_0)$ must not change from the state of the system $S+M$ at $\tau=\tau(t_0)-\epsilon$.

As a result of this condition, in order for selective measurement to be possible, in the state (\ref{eq:No1}) at $\tau=\tau(t_0)-\epsilon$, up to using the wide-sense unitary transformations, one cannot make the state of the system $S$ be Eq.(\ref{eq:No3a}) by separating the systems $S$ and $M$ from each other.

This is equivalent to saying that when selective measurement is possible, at $\tau=\tau(t_0)-\epsilon$, the state $\widehat{\rho}$ (or, equivalently $\widehat{\varrho}$, due to the argument given in the paragraph next to Eq.(\ref{eq:RHO})) of the system $S+M$ does not belong to the same wide-sense unitary equivalence class of the state of the system $S+M$ (we denote this by $\widehat{\rho}^\prime$ or, described in the other way, $\widehat{\varrho}^\prime$) for which separating the systems $S$ and $M$ from each other reduces the state of the system $S$ to Eq.(\ref{eq:No3a}).

This statement can be shown by considering its contraposition in the direct description of the non-selective measurement occurrence by using the density matrix $\widehat{\varrho}$.
First, the state after the selective measurement and the state after the non-selective measurement are {\it not} unitarily equivalent to each other in the narrow sense and belong to {\it different} narrow-sense unitary equivalence classes.
This fact is seen by applying the argument of von Neumann's infinite regression of measuring systems.
(The mathematical proof of the argument of {\it infinite regression} is given in Appendix B.)
Second, when we consider the negation of the consequent of the statement, the condition B2 implies that, at $\tau=\tau(t_0)$, the state $\widehat{\varrho}^\prime$ belongs to the narrow-sense unitary equivalence class of the state after the non-selective measurement.
Furthermore, due to the assumption A2 the state $\widehat{\varrho}$ of the system $S+M$ at $\tau=\tau(t_0)$ also belongs to this narrow-sense unitary equivalence class, which is different from the narrow-sense unitary equivalence class of the state after the selective measurement.
Thus, the selective measurement is impossible: this is the negation of the assumption of the statement.

In order for the consequent of the statement to hold, the non-selective measurement process of the system $S$ (i.e., step (ii) of the scheme) must not be closed in the system $S$ (refer to the left figure in Fig.6).

Next, we {\it assume} that an $M$ playing such a role exists.
With this assumption, the statistical description of the non-selective measurement occurrence by using the density matrix $\widehat{\rho}$ gives two consequences.
First, the amount of transferred entropy ${\sigma}_{M\to {{S}}}$ is $-1$.
This quantity is attributed to the factor $e^{\Delta_r}$ of Eq.(\ref{eq:int0}) corresponding to the second term of the right-hand side of Eq.(\ref{eq:world}), that is, to the term for the non-selective measurement.
Second, the amount of transferred entropy ${\sigma}_{{{S}}\to M}$ is $1$.
This quantity is attributed to the factor (precisely, to the statistical weight for the absence of the non-selective measurement in the enlarged statistical ensemble of copies of the system $S+M$) $e^{-1}$ of Eq.(\ref{eq:int0}) corresponding to the first term of the right-hand side of Eq.(\ref{eq:world}), that is, to the term for a two-fold process: the reduced Hamiltonian process and the time-dependent process representing {\it the absence of} the non-selective measurement.
In fact, since it is definite that one of the factors for normalization $e^{\Delta_r(0)}$ acts on the density matrix of ${{S}}$, if the statistical weight $e^{-1}$ is owned by ${{S}}$, namely if ${\sigma}_{M\to {{S}}}=0$ holds, then the non-selective measurement of ${{S}}$ holds as a closed process in the system ${{S}}$.
This means that $M$ does not satisfy the condition B2 and is disqualified (see the left figure in Fig.6).
In another disqualified case, if the statistical weight $e^{-1}$ is shared by both $M$ and ${{S}}$ with finite amounts, namely, if ${\sigma}_{M\to {{S}}}\neq 0, -1$ holds, then this situation contradicts the independence of $M$ and ${{S}}$ as systems conditioned by the time-dependent process in the system $S+M$ representing the absence of the non-selective measurement:
\begin{equation}
\widehat{\rho}(\tau+d\tau)=(1-\delta(\tau)d\tau)\widehat{\rho}(\tau)\;,
\end{equation}
so $M$ does not satisfy the condition B1 (see the right figure in Fig.6).
In Sec. 3, it will be shown that in the case where $\tau_{\rm et}>d\tau_0$ holds changing the total system $S+M$ at $\tau=d\tau_0$ (shown in the left figure in Fig.6) to the total system $S+M$ at $\tau=\tau_{\rm et}$ (shown in Fig.5) requires finite internal work from the system $M$ to the system $S$.

In summary, under the definition of the measuring system $M$ given above, we obtain the results
\begin{eqnarray}
{\sigma}_{M\to {{S}}}&=&-1\;,\label{eq:ResultR}\\
{\sigma}_{{{S}}\to M}&=&1\;.\label{eq:ResultR2}
\end{eqnarray}

Here, we have to add a point.
In the main statement of this paper, we have assumed that in the selective measurements, the combined measured system ${{S}}$ cannot read any event.
However, we cannot exclude the case in which the combined measured system can read events.
Here, we denote this combined measured system by $M$.
We consider this setting to be another type of selective measurement.
In this setting, the entropy production is of course {\it zero} in the sense of both the internal quantity ${\sigma}_{M\to M}$ and the net quantity ${\sigma}_{M}$ (see Fig.7).
So, in the discussion of the next section, selective measurements of such type require no finite work to be done.

\begin{figure}[htbp]
\begin{center}
\includegraphics[width=0.48\hsize,bb=0 0 260 260]{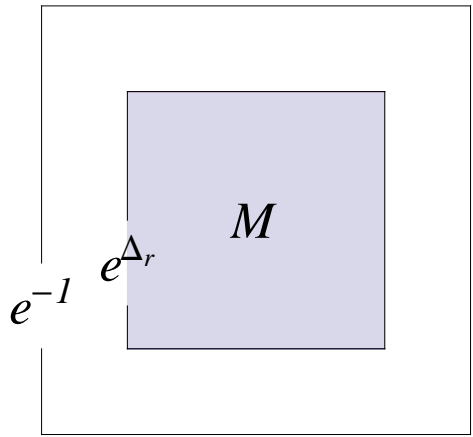}
\end{center}
\caption{
This figure schematically shows the factorization of the normalization factor $e^{-(1-\Delta_r)}$ in Eq.(\ref{eq:int0}) in the case of another type of selective measurement.
Of course, in this selective measurement type, the entropy transfer process does not exist.}
\end{figure}

\section{Thermodynamics of the Measured System}
In general, when the dynamics of a system is described by a Hamiltonian, in the thermodynamic treatment of the system (i.e., for some of the time, the density matrix of the system takes a particular value associated with the Boltzmann-Gibbs distribution), the second law of thermodynamics in all isothermal processes, not only quasi-static processes, starting from a thermal equilibrium state (the final state need not be in thermal equilibrium) can be precisely formularized by the Jarzynski equality\cite{Jarzynski}.
Based on the discussion in Sec. 2, we now consider how the quantum Jarzynski equality\cite{Tasaki1,Mukamel} for the combined measured system $S$ would be modified by including the event reading process.
(When we consider the {\it net} effects of event reading on the total system $S+M$, the quantum Jarzynski equality with measurement takes the same form as that without event reading\cite{Tasaki}.)

In this section, we remove the contraction of the time interval during step (ii) of the measurement scheme (i.e., Eq.(\ref{eq:TAUT})) and invoke the results obtained in Sec. 2.

In the following text, to treat the event reading effect as an additional effect, we use the density matrix $\widehat{\varrho}_{{0}}^S$ of the combined measured system $S$ that corresponds to the density matrix $\widehat{\rho}_{{0}}^S$ defined by Eq.(\ref{eq:Key00}) for $Y=S$ (i.e., having no entropy transfer).
Time evolution of the state $\alpha_0^S$, given by $\alpha_0^S(\widehat{{\cal O}}_S)={\rm tr}_S[\widehat{{\cal O}}_S\widehat{\varrho}_{{0}}^S]$\cite{Araki}, is governed by a Hamiltonian $\widehat{{\cal H}}^S(t)$ (see Eq.(\ref{eq:HS})) that contains the non-selective measurement process but assumes there is {\it no} event reading process by the system $M$.

\subsection{Preliminary}
Now, we consider an isothermal non-equilibrium process in the time interval $[0,t_f]$ switching between an initial thermal equilibrium state at $t=0$ and the final state at $t=t_f$ for a thermodynamic quantum system $S=S_0+A$, based on a thermodynamic treatment of the combined measured system $S$.
During this process, the system $S$ is an isothermal system and is assumed to be externally driven and weakly coupled to a heat bath $B$ that is traced out (refer to footnote $\dagger\dagger$).

We assume that this process contains non-selective measurements (for both the states $\alpha^S$, given by $\alpha^S(\widehat{{\cal O}}_S^{(\star)})={\rm tr}_S[\widehat{{\cal O}}_S^{(\star)}\widehat{\varrho}^S]$, and $\alpha^S_0$) and event readings (for the state $\alpha^S$ only).
These subprocesses work within $t_{{\rm in}}\le t\le t_{{\rm out}}$ and at $t=t_0$ for an arbitrary number of instances $t_{\rm in}$, $t_{\rm out}$, and $t_{0}$.
In this scenario, $0=t_{{\rm ini},{\rm out}}<t_{{\rm ini},0}<t_{1,{\rm in}}< t_{1,{\rm out}}<t_{1,0}<t_{2,{\rm in}}<\ldots<t_{f,{\rm in}}<t_{f,{\rm out}}<t_f=t_{f,0}$.
At $t=t_{{\rm ini},0}$, an energy event reading is performed, and during $[t_{f,{\rm in}},t_{f,0}]$, a selective energy measurement is performed.
During $t_{{\rm ini},0}\le t\le t_f$, the system $A$ is assumed to be in an energy eigenstate.
We denote the total set of the double of instances $(t_{{\rm ini},{\rm out}},t_{{\rm ini},{0}})$ and the triples of instances $(t_{\rm in},t_{\rm out},t_{0})$ by ${\mathfrak S}$.

For $0\le t\le t_{{\rm ini},0}$ and $t_{i,0}\le t\le t_{i+1,{\rm in}}$, we specify two settings.
First, we decouple the systems $S_0$ and $A$ by letting their interaction be zero.
Second, to give a finite quantum uncertainty to the continuous superselection rule $p$ of the measurement apparatus $A$, we localize the pointer position $q$ of the measurement apparatus $A$ by using a valley potential of $q$.

During each subprocess in $t_{i,{\rm out}}\le t<t_{i,0}$ for $i={\rm ini},1,2,\ldots,f$, the state $\alpha^S_{{0}}$ of the combined measured system $S$ is not disturbed; thus, the time evolution of the state $\alpha^S_{{0}}$ is described by using the Hamiltonian
\begin{equation}
{\widehat{{\cal H}}^{{S}}(t)\equiv \widehat{{\cal H}}_{\rm tot}(t)-\widehat{{\cal H}}^M_{\rm kin}-\sum_{i\in {\mathfrak S}}\widehat{{\cal H}}_{{\rm int},t_{i,0}>t\ge t_{i,{\rm out}}}\;.}\label{eq:HS}
\end{equation}

In the rest of this section, we describe the time evolution of the system $S$ in the {\it Heisenberg picture}.

We make three preliminary definitions.

First, we denote the Heisenberg Hamiltonian of the system $S$ whose form in the Schr$\ddot{{\rm o}}$dinger picture is given in Eq.(\ref{eq:HS}) by $\widehat{{\cal H}}^S_H(t)$ or $\widehat{{\cal H}}_{H,\lambda}^{{S}}$.
Here, we set $\widehat{{\cal H}}_H^S(0)=\widehat{{\cal H}}^S(0)$.

Second, for a natural number $N$, we set the time interval $\delta t\equiv t_f/N$ and set time $t_n\equiv n \delta t$ where $n=0,1,\ldots,N$.
Within the time interval $[0,t_f]$, we denote the system's time dependent control parameter by $\lambda$, the time-ordering product by ${\cal T}_>$, the partition function by $Z_\lambda\equiv {\rm{tr}}_{{S}}[e^{-\beta \widehat{{\cal H}}_{H,\lambda}^{{S}}}]$, and the canonical distribution by $\widehat{\varrho}_{\rm can,\lambda}^S\equiv e^{-\beta \widehat{{\cal H}}_{H,\lambda}^S}/Z_\lambda$, where we set $\beta\equiv 1/(k_BT)$.

Third, we define the Helmholtz free energy by $F_\lambda\equiv -\beta^{-1}\ln Z_\lambda$.
Then, the Helmholtz free energy difference between the canonical ensembles of the hypothetical equilibrium system at $t=t_f$ and the equilibrium system at $t=0$ is $\Delta F\equiv -\beta^{-1}\ln (Z_{\lambda_{t_f}}/Z_{\lambda_0})$.

\subsection{Quantum Jarzynski equality}

Since the quantum mechanics of the combined measured system $S$ is described by a time-dependent Hamiltonian $\widehat{{\cal H}}^S_{H,\lambda}$, the quantum Jarzynski equality with no event reading is\cite{Work}
\begin{eqnarray}
&&\overline{\exp(-\beta W)}^0\nonumber\\
&=&{\rm tr}_{{S}}\Biggl[\lim_{N\to \infty}{\cal T}_>\Biggl\{\prod_{n=0}^{N-1}e^{-\beta (\widehat{{\cal H}}_{H,\lambda_{t_{n+1}}}^{{S}}-\widehat{{\cal H}}_{H,\lambda_{t_n}}^{{S}})}\Biggr\}\frac{e^{-\beta \widehat{{\cal H}}_{\lambda_0}^{{S}}}}{Z_{\lambda_0}}\Biggr]\label{eq:MGF1}\\
&=&{\rm tr}_{{S}}\Biggl[\frac{e^{-\beta \widehat{{\cal H}}_{H,\lambda_{t_{f}}}^{{S}}}}{Z_{\lambda_0}}\Biggr]\label{eq:MGF15}\\
&=&\exp(-\beta \Delta F)\label{eq:MGF2}\;.
\end{eqnarray}
Here, the definition of the moment-generating function of quantum work $\overline{\exp(-\beta W)}^0$ and the consequent derivation of Eq.(\ref{eq:MGF1}) will be given in Appendix C.
(Note that $\widehat{{\cal H}}_{\lambda}^S$ contains the non-selective measurement process.)

As shown in Appendix C, two selective measurements of energy for the initial canonical ensemble during $[0,t_{{\rm ini},0}]$ and the final measured state during $[t_{f,{\rm in}},t_f]$ are {\it necessary and sufficient} in the definition of the moment-generating function of work in the quantum regime\footnote{The definition of the quantum work distribution function is still controversial.
For a recent study on this issue assuming no measurement, see Ref.\cite{DPZ}.}\cite{Tasaki1,Work,RCP,TH}.
Here, we consider the initial selective energy measurement.
Since the quantum canonical ensemble of the system $S$ is a state after non-selective energy measurement of the system $S$\footnote{If we invoke the eigenstate thermalization hypothesis (ETH) or its alternative\cite{ETH,ETH2,Thermalization}, we can model the procedure to obtain the canonical ensemble of the system $S$ in two steps:
(I) decoherence of a pure state whose state vector belongs to the substate space associated with a narrow energy span in the energy eigenbasis (i.e., non-selective energy measurement) of the isolated total system $S+B$, and (II) the partial trace of the state of the total system $S+B$ over the heat bath $B$.
Step (I) would give rise to the microcanonical ensemble of the isolated system $S+B$ according to the ETH or its alternative.
Step (II) would give rise to the canonical ensemble of the system $S$ in the thermodynamic limit of the heat bath $B$.
We assume the heat bath $B$ in the setup only for this initial thermalization at $t=0$.}, the set ${\mathfrak S}\backslash \{0\}$ contains only the instance $t=t_{{\rm ini},0}\gtrsim 0$ for the energy event reading.
In the process for the energy event reading (steps (iii) and (iv) in our measurement scheme), by changing $\widehat{{\mathscr{O}}}^{S_0}\otimes \widehat{1}^A\otimes \widehat{1}^M$ in $\widehat{{\cal H}}_{{\rm fb}}$ to $\widehat{{\cal H}}^S\otimes \widehat{1}^M$ during step (iii), we can invoke the results in Sec. 2.
In the following formulae, the factor that gives the work $k_BT$ (in the sense of our main statement) required for this initial energy event reading is contained as the element $t_{{\rm ini},0}$ in the set ${\mathfrak S}$.
During $[0,t_{{\rm ini},0})$, the state of the combined measured system $S$ is not disturbed.

Now, to take into account the event reading effect, in Eq.(\ref{eq:MGF1}), we insert the numbers $e^{{\sigma}_{M\to S,i}}=e^{-1}$ (here, we follow Eq.(\ref{eq:Key})) and the superoperators $\widetilde{\Delta_r}_{t_i,{0}}$ for $i\in{\mathfrak S}$ into the trace while keeping the time-ordering product ${\cal T}_>$.
Each number $e^{\sigma_{M\to S,i}}$ transforms the observable of the system $S$  before and after time $t_{{\rm et}}$, before the event reading by the system $M$ subsequent to the $i$-th non-selective measurement of the system $S$ at $t_{\rm out}$:
\begin{equation}
t_{\rm out}\le t_{\rm et}<t_0\;.
\end{equation}
Each superoperator $\widetilde{\Delta_r}_{t_i,{0}}$ gives the ensemble average of the selective measurement results, by the trace, at the $i$-th event reading\cite{Tasaki}.
In the following, we assume that the density matrix $\widehat{\varrho}^{{S}}_{{0}}$ is non-degenerate (i.e., invertible) at any definition time within $[0,t_f]$.
The quantum Jarzynski equality for the system ${{S}}$ with event readings then takes the form
\begin{eqnarray}
&&\overline{\exp\Biggl(-\beta W+\sum_{i\in{\mathfrak S}} 1\Biggr)}\nonumber\\
&=&{\rm tr}_{{S}}\Biggl[\lim_{N\to\infty}{\cal T}_>\Biggl\{\Biggl(\prod_{n=0}^{N-1}e^{-\beta (\widehat{{\cal H}}^{{S}}_{H,\lambda_{t_{n+1}}}-\widehat{{\cal H}}^{{S}}_{H,\lambda_{t_n}})}\Biggr)\Biggl(\prod_{j\in {\mathfrak S}}\widetilde{\Delta_r}_{t_j,{0}}\Biggr)\Biggl(\prod_{j\in{\mathfrak S}}e^{{\sigma}_{M\to S}}
\Biggr)\Biggr\}\nonumber\\
&&\Biggl(\prod_{i\in{\mathfrak S}} e^{\sigma_{S\to M}}\Biggr)\frac{e^{-\beta \widehat{{\cal H}}^{{S}}_{\lambda_0}}}{Z_{\lambda_0}}\Biggr]\label{eq:JarA}\\
&=&{\rm tr}_{{S}}\Biggl[\frac{e^{-\beta \widehat{{\cal H}}^{{S}}_{H,\lambda_{t_{f}}}}}{Z_{\lambda_0}}\Biggr]\label{eq:JarB}\\
&=&\exp(-\beta \Delta F)\;.\label{eq:Jar}
\end{eqnarray}
To derive Eq.(\ref{eq:JarB}) from Eq.(\ref{eq:JarA}), we use
\begin{eqnarray}
&&
{\cal T}_>\Bigl\{e^{-\beta(\widehat{{\cal H}}^S_{H,\lambda_{t_{n+1}}}-\widehat{{\cal H}}^S_{H,\lambda_{t_n}})}\widetilde{\Delta_r}_{t}e^{-\beta \widehat{{\cal H}}_{H,\lambda_{t_n}}^S}\Bigr\}
\nonumber\\
&\equiv&e^{-\beta\widehat{{\cal H}}^S_{H,\lambda_{t_{n+1}}}}\widetilde{\Delta_r}_{t}(\widehat{1}^S)\\
&=&e^{-\beta\widehat{{\cal H}}^S_{H,\lambda_{t_{n+1}}}}
\end{eqnarray}
for the non-negative integer $n$ which satisfies ${t_{n}}\le t<{t_{n+1}}$.

The second term in the exponential of the left-hand side of Eq.(\ref{eq:JarA}) comes from the entropy production accompanying the selective measurement; it is to be contained in the second part of the entropy production formula $\overline{\sigma}\equiv \beta\overline{W}^0-\beta \Delta F$ where
\begin{equation}
\overline{W}^0\equiv {\rm tr}_{{S}}\Bigl[\widehat{{\cal H}}^S_{H,\lambda_{t_f}}\widehat{\varrho}_{0}^S(0)\Bigr]-{\rm tr}_{{S}}\Bigl[\widehat{{\cal H}}^S_{\lambda_0}\widehat{\varrho}_{0}^S(0)\Bigr]\label{eq:Wreg}
\end{equation}
is the total average work in the switching process (see Appendix C) and $-\beta\Delta F=\Delta \ln Z$.
This term cannot be derived from the time evolution
without event reading and this term arises from selective measurement as the result of the redefining of observables (see Eq.(\ref{eq:Key})).

Here, we make a significant note.
Eq.(\ref{eq:MGF1}) is the average not of an observable defined at a single time but, in our setting of an isothermal process starting from a thermal equilibrium state, of an operator $\widehat{{\cal O}}_{{{S}},{\cal I}}$ defined {\it over a finite time interval ${\cal I}\equiv [0,t_f]$}.
This is the unique operator, up to multiplication by conservative observables, in which, for an arbitrary intermediate time $t^{(0)}$ in ${\cal I}$, the expression
\begin{equation}
\overline{\widehat{{\cal O}}_{{{S}},{\cal I}}}^0={\rm tr}_{{S}}[\widehat{{\cal O}}^\prime_{{{S}},t_f\ge t \ge t^{(0)}}(t^{(0)})\widehat{\varrho}_{{0}}^{{S}}(t^{(0)})]\;,\label{eq:OO}
\end{equation}
where
\begin{eqnarray}
\widehat{{\cal O}}^\prime_{{S},t_f\ge t \ge t^{(0)}}(t^{(0)})
\equiv (\widehat{{\cal O}}_{{S},t_f\ge t \ge t^{(0)}})(\widehat{\varrho}^{{S}}_{\rm can}(t^{(0)}))(\widehat{\varrho}^{{{S}}}_{{0}}(t^{(0)}))^{-1}\frac{Z_\lambda}{Z_{\lambda_0}}\;,\label{eq:Picture}
\end{eqnarray}
always holds as a renewing of the definition time of the density matrix $\widehat{\varrho}_0^S$ at which $\widehat{\varrho}_0^S$ is fixed from $0$ to $t^{(0)}$ (see Appendix C).
This is a proper feature of work $W$.
Then, Eq.(\ref{eq:Key}) (which is applicable only to the observables $\widehat{{\cal O}}_Y$ defined at time $t$ after entropy is transferred from the system $M$ to the system $S$: $t\ge t_{\rm et}$) can be used to define $\overline{\widehat{{\cal O}}_{{{S}},{\cal I}}}$ (i.e., $\widehat{{\cal O}}_{Y}(t^{(0)})=\widehat{{\cal O}}^\prime_{{{S}},t_f\ge t \ge t^{(0)}}(t^{(0)})$ for $t^{(0)}\ge t_{\rm et}$ in Eq.(\ref{eq:Key})) by inserting the numbers $e^{{\sigma}_{M\to S}}=e^{-1}$ and the superoperators $\widetilde{\Delta_r}$ into the trace (\ref{eq:OO}) as in Eq.(\ref{eq:JarA}).
This is because, at each $t^{(0)}$ in the time interval $0\le t^{(0)}< t_{\rm et}$, the average object in this $\overline{\widehat{{\cal O}}_{{{S}},{\cal I}}}$ does not contain any non-conservative observable {\it defined before} $t^{(0)}$ that would prevent the renewing of the definition time of the density matrix $\widehat{\varrho}_0^S$ in $\overline{\widehat{{\cal O}}_{{{S}},{\cal I}}}$ from $0$ at the definition time of this observable.

Of course, when we consider the case of the average $\la{\widehat{{{\cal O}}}_{{S}}}\ra$ of an observable $\widehat{{\cal O}}_{{S}}$ defined at a single time, the effect of entropy production attributed to selective measurements does not appear in the result of the averaging.
However, to let $\overline{\widehat{{\cal O}}_{{{S}},{\cal I}}}$, which is modified by the event readings, be a quantum statistical average using a well-defined statistical operator (density matrix) $\widehat{\varrho}_0^S$, we must redefine work by a shift of $|{\mathfrak S}|/\beta$ (see Eq.(\ref{eq:Result})).
So, though the average of work $\overline{W}^0$ is not the type of Eq.(\ref{eq:OO}), it is exceptionally modified by event readings.

\subsection{Thermodynamic inequality}

As a consequence of the result obtained in the last subsection, combining the modified quantum Jarzynski equality in Eq.(\ref{eq:Jar}) with the Jensen inequality $\overline{\exp(x)}\ge \exp(\overline{x})$ for $x=-\beta W$, we arrive at the thermodynamic inequality in the combined measured system $S$
\begin{equation}
\Delta F+k_BT \sum_{i\in{\mathfrak S}}1 \le \overline{W}\;.
\end{equation}
This inequality means that for a single selective measurement, the production of negative entropy for the combined measured system, which can be directly transformed into an amount of Helmholtz free energy of $k_BT$ for the system, is needed.

To satisfy this inequality consistently, with the redefinitions of the observables of the combined measured system ${{S}}$ in $\overline{\exp(-\beta W)}$ by selective measurements, the quantity of work
\begin{equation}
W_{{\rm e.r.}}=k_BT\sum_{i\in {\mathfrak S}}1\label{eq:Result}
\end{equation}
needs be added to the regular quantity of work in the absence of event reading (e.r.) $\overline{W}^0$ in Eq.(\ref{eq:Wreg}): $\overline{W}=\overline{W}^0+W_{{\rm e.r.}}$.
Thus, the measuring system $M$ is required to do an amount of work of $k_BT=k_BT\sigma_{S\to M}$ in the event reading process, which is independent of the fineness of the measurement for a non-trivial initial state, in every selective measurement of the combined measured system ${{S}}$.
(For the trivial initial state, the combined measured system ${{S}}$ completes the selective measurement, and the measuring system $M$ does not exist.)
This statement is the main result of this paper.
Here, we note again that neither generation of heat by the combined measured system $S$ nor absorption of heat by the measuring system $M$ accompanies the entropy transfer $\sigma_{S\to M}$ (see footnote $\S$), and the net amount of work in the total system $S+M$ for measurement is zero.

\section{Summary and Discussion}
In this paper, by using the density matrix, we have studied the single projective quantum measurement of a discrete or discretized continuous observable of a quantum system.
Here, {\it quantum measurement} is a selective measurement defined by non-selective measurement plus its subsequent event reading.
The combined measured system $S$ of the measured system $S_0$ and the measurement apparatus $A$ is the system in which the non-selective measurement part acts.

We defined the concept of a {\it measuring system} $M$ by two conditions.
Firstly, this system $M$ is independent from the combined measured system ${{S}}$ (in which case we call it a {\it type I} selective measurement) or inseparable from the combined measured system ${{S}}$ (in which case we call it a {\it type II} selective measurement).
Secondly, when we can separate the systems $S$ and $M$ from each other without change of the wide-sense unitary equivalence class of the state of the system $S$ from that obtained by the partial trace of the system $M$, the selective measurement cannot be completed.

We analyzed the solution of the von Neumann equation for the non-selective measurement process treated as a cut-off inhomogeneous one-time Poisson process by invoking von Neumann's argument of infinite regression of measuring systems in state reduction.
As a result, we found that, in a type I (type II) selective measurement, there is an entropy transfer of minus one unit (zero) from the measuring system $M$ to the combined measured system ${{S}}$ due to the event reading process.
(Neither generation of heat by the combined measured system $S$ nor absorption of heat by the measuring system $M$ accompanies this entropy transfer.)
In a type I selective measurement, entropy transfer indicates the non-divisibility of the state of the total system into those of the subsystems $S$ and $M$ when holding the unitary equivalence classes of the states of the subsystems $S$ and $M$ constant (note footnote $+$), and the transferred entropy originates in the reduction of our knowledge about the total system.
This reduction is due to the averaging operation in the statistical treatment of one non-selective measurement process as a cut-off one-time Poisson process.
In the total system, this knowledge that is an analogue of information is lost by the measuring system $M$, and the lost knowledge by the diagonal part of the density matrix is gained by the combined measured system $S$.

For this fact, in a type I selective measurement, work is required to be done by the measuring system $M$ to the combined measured system $S$.
Indeed, in the thermodynamic treatment of the combined measured system ${{S}}$, from the modification of the quantum Jarzynski equality, we found that, in type I selective measurement, this transferred entropy can be directly transformed into an amount of Helmholtz free energy of $k_BT$ of the combined measured system ${{S}}$.
Equivalently, an internal work $k_BT$ or $0$
\begin{equation}
W_{{\rm e.r.}}:\xymatrix{{{{M}}}\ar@[black]@(ul,dl)[]|{0}\ar@[black][rr]^{k_BT}&&{{S}}}\ \ ({{S}}\neq M)\label{eq:SM}
\end{equation}
is required to be done by the system $M$ for one type I or type II selective quantum measurement.
For type I selective measurement, the internal work $k_BT$ is required in the event reading process.
So, this result is independent of the fineness of the measurement for a non-trivial initial state.
In both types of measurements, the net amount of entropy production is zero and no net amount of work is required in the total system $S+M$.

Due to these results, we can conclude that, if a measuring system matching our definition exists, in type I selective quantum measurement, the question posed at the beginning of this paper would be answered partially in the affirmative: while type II selective quantum measurement is {\it not} in itself a physical process, type I selective quantum measurement {\it is} in itself a physical process.

We close this paper with three comparisons of our results with those of other theories.

First, our result about the modification of work by quantum measurement cannot be applied to the non-selective measurement in the program of decoherence\cite{Decoherence,Decoherence2a}.
Actually, if there is no subsequent event reading, the entropy transfer ${\sigma}_{\bar{Y}\to Y}$, which leads to the work modification, is always zero for an arbitrary subsystem $Y$ of the total system ${{S}}+M$.

Second, we comment on the connection of our result to Refs.\cite{SU1,SU2}.
In Ref.\cite{SU2}, the change of von Neumann entropy by a non-selective quantum measurement, as in Eq.(\ref{eq:world}), in the combined system of the measured system ${{S}}$, the memory system ${\cal M}$ and the heat baths $B=\{B_m\}$ (${\cal M}$ and $B$ are initially in canonical distributions and the state space of ${\cal M}$ is a direct sum of the subspaces corresponding to the measurement outcomes as in Eq.(\ref{eq:Str})) is calculated and the lower-bound for the work $W_{\rm meas}$ required to make a non-selective measurement is derived.
When we call the system $M$ the memory system ${\cal M}$, our result of internal energy transfer $W_{{\rm e.r.}}$ in type I selective measurement (but, not in type II selective measurement) would modify this lower-bound for the work $W_{\rm meas}$ by adding $W_{{\rm e.r.}}$: however, of course, it would not modify the work $W_{\rm eras}$ required to erase the memory.
The lower-bound of $W_{\rm meas}+W_{\rm eras}$ is another result of Ref.\cite{SU2}, such that originally 
\begin{equation}
W_{\rm meas}+W_{\rm eras}\ge k_BT I\label{eq:SUII}
\end{equation}
holds for the QC-mutual information $I$\cite{QC1,QC2} (which satisfies $0\le I\le H(\{p_n\})$ for the Shannon entropy $H(\{p_n\})$ of probabilities $p_n$ for the measurement outcomes $x_n$\cite{SU1}) associated with the non-selective measurement of the system $S$\cite{SU2}.
Here, a non-selective measurement of the system $S$ corresponds to an error-free measurement of the system $S+{\cal M}$ in Ref.\cite{SU2}.

Aside from the measurement process, the thermodynamics of information consists of two basic processes.

The first process is the {\it feedback process} to convert the QC-mutual information $I$, gained by the measurement, to a process with extra work $W_{\rm meas}+\Delta W_{\rm gain}$ for the work\cite{SU1}
\begin{equation}
\Delta W_{\rm gain}\ge -k_BTI\;.\label{eq:SUI}
\end{equation}
For instance, this is, in a step-by-step description,
\begin{eqnarray}
&&
\widehat{\varrho}^{S}_{{\rm can}}\otimes\widehat{\varrho}^{{\cal M}}_{0,{\rm can}}\otimes \widehat{\varrho}^{B}_{\rm can}
\nonumber\\
&\longrightarrow &\Bigl(\widehat{U}_{\rm i}\Bigl(\widehat{\varrho}^{S}_{{\rm can}}\otimes\widehat{\varrho}^{B^{(1)}}_{\rm can}\Bigr)\widehat{U}_{\rm i}^{-1}\Bigr)\otimes \widehat{\varrho}^{{\cal M}}_{0,{\rm can}}\otimes \widehat{\varrho}^{B^{(2)}}_{\rm can}\\
&\longrightarrow& \sum_{n_0}p_{n_0}\widehat{\varrho}^{S,B^{(1)}}_{n_0}\otimes \widehat{\varrho}^{{\cal M}}_{0,{\rm can}}\otimes \widehat{\varrho}^{B^{(2)}}_{\rm can}\\
&\longrightarrow& \sum_{n_0}p_{n_0}\widehat{\varrho}^{S,B^{(1)}}_{n_0}\otimes\widehat{\varrho}^{{\cal M},B^{(2)}}_{n_0}\\
&\longrightarrow& \sum_{n_0}p_{n_0}\Bigl(\widehat{U}_{n_0}\widehat{\varrho}^{S,B^{(1)}}_{n_0}\widehat{U}_{n_0}^{-1}\Bigr)\otimes\widehat{\varrho}^{{\cal M},B^{(2)}}_{n_0}\\
&\longrightarrow& \sum_{n_0}p_{n_0}\Bigl(\widehat{U}_{\rm f}\Bigl(\Bigl(\widehat{U}_{n_0}\widehat{\varrho}^{S,B^{(1)}}_{n_0}\widehat{U}_{n_0}^{-1}\Bigr)\otimes\widehat{\varrho}^{{\cal M},B^{(2)}}_{n_0}\Bigr)\widehat{U}_{\rm f}^{-1}\Bigr)\label{eq:feedback}
\end{eqnarray}
for the standard memory state $n=0$, disjoint subsets $B^{(1)}$ and $B^{(2)}$ of $B=B^{(1)}\cup B^{(2)}$, a unitary operator $\widehat{U}_{\rm i}$ and feedback unitary operators $\widehat{U}_n$ acting on the state space of the system $S+B^{(1)}$, equilibration unitary operator $\widehat{U}_{\rm f}$ acting on the state space of the system $S+{\cal M}+B$, and the density matrix linear component $\widehat{\varrho}^{S,B^{(1)}}_{n_0}\otimes \widehat{\varrho}^{{\cal M}}_{0,{\rm can}}\otimes \widehat{\varrho}_{\rm can}^{B^{(2)}}$ of the system $S+{\cal M}+B$, after a non-selective measurement of the system $S$, with outcome $n_0$.
Here, the support of the memory system part of $\widehat{\varrho}^{{\cal M},B^{(2)}}_{n_0}$ belongs to ${\cal V}_{n_0}^{\cal M}$ with unit probability.

The second process, the unitary {\it erasure process} of memory, is
\begin{eqnarray}
\sum_{n_0}p_{n_0}\widehat{\varrho}^{{\cal M}}_{n_0,{\rm can}}\otimes \widehat{\varrho}_{\rm can}^{B^{(2)}}\longrightarrow \widehat{\varrho}^{\prime{\cal M},B^{(2)}}_{0}\;,\label{eq:erasure}
\end{eqnarray}
where the systems $S$ and $B^{(1)}$ are traced out and the support of the memory system part of $\widehat{\varrho}^{\prime{\cal M},B^{(2)}}_0$ belongs to ${\cal V}_0^{\cal M}$ with unit probability, and requires work $W_{\rm eras}$.

Both processes are {\it physical} and can be treated thermodynamically\cite{SU1,SU2}.
While the feedback process has the gain (\ref{eq:SUI}), the total processes (\ref{eq:feedback}) and (\ref{eq:erasure}) require the work (\ref{eq:SUII}).
Thus, the second law of thermodynamics, as Planck's and Kelvin's principles, holds for the thermodynamics of information.
From these arguments and our result, the information quantities (of which the mutual information is the key quantity) and the entropy transfer enable us to treat projective quantum measurements in information processing thermodynamically\cite{SU1,SU2,SagawaPTP,Sagawa}.

Finally, we compare our result with a recent result, derived in Ref.\cite{Reeb}, about the net fundamental energy cost, $W^{{\cal M}}_{\rm proj}$, of a projective measurement and the erasure of memory required by the memory system ${\cal M}$ (here, we omit the heat bath) after tracing out (i.e., averaging out) the system $S$.
The result is
\begin{equation}
W^{{\cal M}}_{\rm proj}=k_BTH(\{p_n\})\;,\label{eq:Reeb}
\end{equation}
which depends on the fineness of the initial superposition of the system $S$.
At first glance, this result contradicts our result (\ref{eq:Result}).
However, the result (\ref{eq:Reeb}) is concerned with the net energy cost due to the state change of the memory system ${\cal M}$ after tracing out the system $S$.
In contrast, our result (\ref{eq:Result}) is concerned with the internal work $k_BT$ with no net work due to the event reading (and not for the program of decoherence).
So, these two results do not contradict each other.
It is worth noting that if one does not trace out the system $S$, then the work formula that corresponds to Eq.(\ref{eq:Reeb}) is the main result (\ref{eq:SUII}) from Ref.\cite{SU2}, as has been explained above.

\ack

The author thanks Dr. M. M. Sano for advice on improving the manuscript and the anonymous referees for their suggestions and constructive criticism.

\begin{appendix}
\section{Grounds for Assumption A1}
In this appendix, we explain the basis for assumption A1 (Sec. 2.2) of quantum mechanical equivalence between the direct description and the statistical description of non-selective measurement occurrence.

We denote by $X$ a pure state of a given system and introduce two natural numbers $M$ and $N$ with $M,N\to \infty$ in the limit.

In the concept of statistical ensemble used in the direct (i.e., conventional) description, when we {\it reduce our knowledge (r.o.k.)} about the state of the system $X$, a pure ensemble changes to a mixture:
\begin{equation}
{{\rm r.o.k.}}:[X]\longrightarrow [X_1,X_2,\ldots,X_N]\;.\label{eq:Des1}
\end{equation}

In contrast, in the concept of an enlarged statistical ensemble used in the statistical description, we {\it reduce our knowledge} about the occurrence time of one non-selective measurement in the system $X$.
Then, when we refer to the mutually exclusive mixture after non-selective measurement as the {\it original mixture}, a `{\it pure ensemble of original mixture}' changes to a `{\it mixture of original mixtures}':
\begin{eqnarray}
{{\rm r.o.k.}}:[X_1,X_2,\ldots,X_N]&{\longrightarrow}& \Bigl[\Bigl[X_1^{(1)},X_2^{(1)},\ldots, X_N^{(1)}\Bigr],\ldots\nonumber \\
&&\ldots,\Bigl[X_1^{(M)},X_2^{(M)},\ldots,X_N^{(M)}\Bigr]\Bigr]\;.\label{eq:Des2}
\end{eqnarray}
This reduction of our knowledge is due to the averaging operation of the non-selective occurrence time.

 Eq.(\ref{eq:Des1}) is stated in the conventional theory of statistical ensemble.
Of course, the pair of the state space and the space of the observables is common to both sides.

Now, we note that, in Eq.(\ref{eq:Des2}), the right-hand side is the `{\it mixture of original mixtures}'.
It applies the idea of reduction of a pure ensemble to a mixture, due to the reduction of our knowledge, to the idea of reduction of an original mixture, in the same way as in Eq.(\ref{eq:Des1}).
So, for both sides of Eq.(\ref{eq:Des2}), we assume that the pair of the state space and the space of the observables is shared in common.
This is the assumption A1 made in Sec. 2.2.

Here, we add a note about the enlarged ensemble.

If and only if the enlarged ensemble is {\it not} a pure ensemble just after an event reading (i.e., if and only if the reduction of our knowledge about the non-selective measurement occurrence in Eq.(\ref{eq:Des2}) does {\it not} lose its validity just after an event reading), then this enlarged ensemble can describe  the occurrence of non-selective measurement statistically {\it only once}.
So, to describe non-selective measurement occurrence statistically and repeatedly, the enlarged ensemble must be a pure ensemble, with state equivalent to the state of the original ensemble, just after every event reading.

\section{von Neumann's Infinite Regression}
Following Ref.\cite{NGthm1}, in this appendix, we mathematically formulate the statement of {\it von Neumann's infinite regression of measuring systems} $M$: {\it the event reading process of the system $M$ cannot be realized by a unitary transformation in the system ${S}+M$ after the non-selective measurement of the system ${S}=S_0+A$ except for the trivial case in which the initial state of the system ${S}$ is an eigenstate.}

The notation of this appendix is the same as that of Sec. 2.

A generic unitary transformation of the density matrix of the system $S_0+M$
after the non-selective measurement of ${S}$ can be written as
\begin{eqnarray}
&&
\sum_n|c_n|^2|x_n,{\mathfrak A}_0\ra|{\mathfrak M}_r\ra\la {\mathfrak M}_r|\la x_n,{\mathfrak A}_0|
\nonumber\\
&&
\to\widehat{\varrho}_{r}^{(1)}=\sum_n|c_n|^2|x_n,{\mathfrak A}_0\ra|{\mathfrak M}_{(rn)}\ra\la {\mathfrak M}_{(rn)}|\la x_n,{\mathfrak A}_0|\label{eq:Phir}\;.
\end{eqnarray}
Note that $\widehat{\varrho}_{r}^{(1)}$ contains one or several indices of $x$.

On the other hand, by admitting an initial state mixture of the system $M$, the form of the density matrix of the system ${S}+M$ after the selective measurement of ${S}+M$ is
\begin{eqnarray}
\widehat{\varrho}_{m}^{(2)}=\sum_r \chi_r^{(m)}|x_m,{\mathfrak A}_0\ra|{\mathfrak M}_{(rm)}\ra\la {\mathfrak M}_{(rm)}|\la x_m,{\mathfrak A}_0|\;.
\end{eqnarray}
Note that $\widehat{\varrho}_{m}^{(2)}$ contains only one index of $x$.

Here, due to the orthonormality of the states $|x_n,{\mathfrak A}_0\ra$, the orthonormality of the states $|{\mathfrak M}_r\ra$ and the unitarity of the transformation in Eq.(\ref{eq:Phir}), the orthonormality of the states $|x_m,{\mathfrak A}_0\ra|{\mathfrak M}_{(rm)}\ra$ of the system ${S}+M$ follows:
\begin{equation}
\la x_m,{\mathfrak A}_0|x_n,{\mathfrak A}_0\ra\la {\mathfrak M}_{(rm)}|{\mathfrak M}_{(sn)}\ra=\delta_{mn}\delta_{rs}\;.\label{eq:Mrm}
\end{equation}

Now, we assume the statement of the von Neumann infinite regression of measuring systems $M$.
This means that the density matrix $\widehat{\varrho}_{m}^{(2)}$ is given by a linear combination (i.e., a mixture with respect to the system $M$) of $\widehat{\varrho}_{r}^{(1)}$
\begin{equation}
\widehat{\varrho}_{m}^{(2)}=\sum_ru_r\widehat{\varrho}_{r}^{(1)}\;.\label{eq:Psimk}
\end{equation}
Then, due to the orthonormality condition, Eq.(\ref{eq:Mrm}), the following relations between the $c$-number coefficients of the states $|x_m,{\mathfrak A}_0\ra|{\mathfrak M}_{(rm)}\ra$ on both sides of Eq.(\ref{eq:Psimk}) must hold
\begin{equation}
u_r|c_n|^2=\delta_{mn}\chi_r^{(m)}\;.
\end{equation}
However, this equation has no solution when more than one $c_n$ is non-zero.
So, this assumption holds in only the trivial case such that the initial state of ${S}$ is an eigenstate of the observable.

Since we assume the initial state of ${S}$ to be a non-trivial one (namely, at least two of $c_n$ are non-zero), this statement is the von Neumann infinite regression of measuring systems $M$ used in the main text.

\section{Quantum Work}

We present this appendix to discuss quantum work.

\subsection{Two energy measurement approach}
In this subsection, we explain the {\it two energy measurement} approach to defining the quantum work distribution function and the moment-generating function of quantum work, derive Eq.(\ref{eq:MGF1}) by applying this approach, and give a sketch of the proof of Eq.(\ref{eq:Picture}).

In the following, the quantum system is initially in the thermal equilibrium state.

To avoid unnecessary complication in the notation, we omit the control parameter $\lambda$ from the notation.

First, we define the quantum work distribution function in terms of the Schr$\ddot{{\rm o}}$dinger picture variables\cite{ReviewJ,Work,TH}.

We denote by $p(W=E_m(t_f)-E_n(0))$ the probability to obtain the energy eigenvalue $E_n(0)$ with energy eigenstate $|\phi_n(0)\ra$ by energy measurement at $t=0$ and obtain the energy eigenvalue $E_m(t_f)$ with energy eigenstate $|\phi_m(t_f)\ra$ by energy measurement at $t=t_f$, where, during $0<t<t_f$, the system evolves according to $\widehat{U}(t)$, which is the solution of the Schr$\ddot{{\rm o}}$dinger equation $i\hbar \pa \widehat{U}(t)/\pa t=\widehat{{\cal H}}(t)\widehat{U}(t)$ with $\widehat{U}(0)=1$.

This quantity can be written as
\begin{eqnarray}
p(W=E_m(t_f)-E_n(0))=|\la \phi_m(t_f)|\widehat{U}(t_f)\phi_n(0)\ra|^2\frac{e^{-\beta E_n(0)}}{Z_0}
\end{eqnarray}
by using the initial time probability distribution $e^{-\beta E_n(0)}/Z_0$.

Using this equality, the quantum work distribution function is given by
\begin{eqnarray}
p(W)=\sum_{m,n}\delta (W-[E_m(t_f)-E_n(0)])p(W=E_m(t_f)-E_n(0))\;.
\end{eqnarray}

Now, following Ref.\cite{Work}, we define the moment-generating function of quantum work and derive Eq.(\ref{eq:MGF1}):
\begin{eqnarray}
\overline{\exp(-\beta W)}^0&\equiv&\int dW e^{-\beta W}\{p(W)\}_1\\
&=&\int dWe^{-\beta W}\Biggl\{\sum_{m,n}\delta(W-[E_m(t_f)-E_n(0)])\nonumber\\
&&p(W=E_m(t_f)-E_n(0))\Biggr\}_1\\
&=&\Biggl\{\int dWe^{-\beta W}\sum_{m,n}\delta(W-[E_m(t_f)-E_n(0)])\Biggr\}_2\nonumber\\
&&|\la \phi_m(t_f)|\widehat{U}(t_f)\phi_n(0)\ra|^2\frac{e^{-\beta E_n(0)}}{Z_0} \\
&=&\Biggl\{\sum_{m,n}e^{-\beta(E_m(t_f)-E_n(0))}\Biggr\}_2\la \phi_m(t_f)|\widehat{U}(t_f)\phi_n(0)\ra \nonumber \\
&&\la \phi_n(0)|\widehat{U}^\dagger (t_f)\phi_m(t_f)\ra \frac{e^{-\beta E_n(0)}}{Z_0}\\
&=&\sum_{m,n}\biggl<\phi_m(t_f)\biggr|\widehat{U}(t_f)\biggl\{\frac{\phi_n(0)}{Z_0}\biggr\}_3\biggr>\nonumber\\
&&\la\phi_n(0)|\widehat{U}^\dagger (t_f)\{e^{-\beta E_m(t_f)}\phi_m(t_f)\}_4\ra\\
&=&\sum_{m,n}\la \phi_m(t_f)|\widehat{U}(t_f)\{e^{\beta \widehat{{\cal H}}(0)}\widehat{\varrho}_{\rm can}(0)\phi_n(0)\}_3\ra\nonumber\\
&&\la \phi_n(0)|\widehat{U}^\dagger (t_f)\{e^{-\beta \widehat{{\cal H}}(t_f)}\phi_m(t_f)\}_4\ra\\
&=&{\rm tr}\widehat{U}(t_f)e^{\beta \widehat{{\cal H}}(0)}\widehat{\varrho}_{\rm can}(0)\widehat{U}^\dagger (t_f)e^{-\beta \widehat{{\cal H}}(t_f)}\\
&=&{\rm tr}e^{-\beta \widehat{{\cal H}}_H(t_f)}e^{\beta \widehat{{\cal H}}(0)}\widehat{\varrho}_{\rm can}(0)\\
&=&{\rm tr}{\cal T}_>e^{-\beta(\widehat{{\cal H}}_H(t_f)-\widehat{{\cal H}}(0))}\widehat{\varrho}_{\rm can}(0)\label{eq:CEq}\;.
\end{eqnarray}
Here, $\widehat{{\cal H}}_H(t)=\widehat{U}^\dagger(t)\widehat{{\cal H}}(t)\widehat{U}(t)$ with $\widehat{{\cal H}}_H(0)=\widehat{{\cal H}}(0)$ is the Hamiltonian in the Heisenberg picture, and ${\cal T}_>$ refers to the time-ordering product in the Heisenberg picture.

It is worth noting that
\begin{eqnarray}
&&{\rm tr}{\cal T}_>e^{-\beta(\widehat{{\cal H}}_H(t_f)-\widehat{{\cal H}}(0))}\widehat{\varrho}_{\rm can}(0)\nonumber\\
&=&{\rm tr}{\cal T}_>\exp\biggl[-\beta \int_0^{t_f}\frac{d \widehat{{\cal H}}_H(s)}{d s}ds\biggr]\widehat{\varrho}_{\rm can}(0)\\
&=&{\rm tr}{\cal T}_>\exp\biggl[-\beta \int_0^{t_f}\frac{\pa \widehat{{\cal H}}_H(s)}{\pa s}ds\biggr]\widehat{\varrho}_{\rm can}(0)\;.
\end{eqnarray}

From the above definition of $\overline{\exp(-\beta W)}^0$, the average work $\overline{W}^0$ is the difference between the expected values for the energy of the system at the initial and the final times, that is, Eq.(\ref{eq:Wreg}).

Finally, we demonstrate the proof of Eq.(\ref{eq:Picture}) in the case of $N=3$ and $t^{(0)}=t_2$ with no event reading process.

By using $\widehat{U}(0)=1$, we obtain
\begin{eqnarray}
\overline{\exp(-\beta W)}^0
&=&{\rm tr}[\widehat{U}^\dagger(t_3)e^{-\beta \widehat{{\cal H}}(t_3)}\widehat{U}(t_3)
\widehat{U}^\dagger(t_2)e^{\beta \widehat{{\cal H}}(t_2)}\widehat{U}(t_2)\nonumber\\
&&\widehat{U}^\dagger(t_2)e^{-\beta \widehat{{\cal H}}(t_2)}\widehat{U}(t_2)
\widehat{U}^\dagger(t_1)e^{\beta \widehat{{\cal H}}(t_1)}\widehat{U}(t_1)\nonumber\\
&&\widehat{U}^\dagger(t_1)e^{-\beta \widehat{{\cal H}}(t_1)}\widehat{U}(t_1)
\widehat{U}^\dagger(0) e^{\beta \widehat{{\cal H}}(0)}\widehat{U}(0)
\widehat{\varrho}_{\rm can}(0)]\\
&=&{\rm tr}\biggl[\widehat{U}^\dagger(t_3) e^{-\beta \widehat{{\cal H}}(t_3)}\widehat{U}(t_3)
\widehat{U}^\dagger(t_2) e^{\beta \widehat{{\cal H}}(t_2)}\widehat{U}(t_2)\nonumber\\
&&\widehat{U}^\dagger(t_2) e^{-\beta \widehat{{\cal H}}(t_2)}\widehat{U}(t_2)\frac{1}{Z_0}\biggr]\;.\label{eq:Sketch}
\end{eqnarray}

Now, we change the definition time of the density matrix of the system from $t=0$ to $t=t_2$.
We denote the corresponding time evolution operator by $\widehat{U}^\prime (t)$.

Then, since $\widehat{U}^\prime(t_2)=1$ holds, Eq.(\ref{eq:Sketch}) becomes
\begin{eqnarray}
&&{\rm tr}\biggl[\widehat{U}^{\prime \dagger}(t_3) e^{-\beta \widehat{{\cal H}}(t_3)}\widehat{U}^\prime(t_3)
\widehat{U}^{\prime \dagger}(t_2) e^{\beta \widehat{{\cal H}}(t_2)}\widehat{U}^\prime (t_2)
e^{-\beta \widehat{{\cal H}}(t_2)}\frac{1}{Z_0}\biggr]\nonumber\\
&&={\rm tr}\biggl[\widehat{U}^{\prime\dagger}(t_3) e^{-\beta \widehat{{\cal H}}(t_3)}\widehat{U}^\prime(t_3)
\widehat{U}^{\prime\dagger}(t_2)e^{\beta \widehat{{\cal H}}(t_2)}\widehat{U}^\prime (t_2)
\widehat{\varrho}_{\rm can}(t_2)\frac{Z_{t_2}}{Z_0}\biggr]\\
&&={\rm tr}\biggl[e^{-\beta \widehat{{\cal H}}_H^\prime(t_3)}e^{\beta\widehat{{\cal H}}_H^\prime(t_2)} \widehat{\varrho}_{\rm can}(t_2)\frac{Z_{t_2}}{Z_0}\biggr]\;.
\end{eqnarray}
This leads to the statement of Eq.(\ref{eq:Picture}).

\subsection{Recent developments}

In this subsection, we explain via measurement theory recent developments in studies of the definition of quantum work and the quantum Jarzynski equality.

In the following, we assume that the process using the force protocol is unitary (i.e., without intermediate event readings).

First, in the two energy measurement approach explained in App. C.1, the projective energy measurements can be generalized to, in particular, more experimentally practical Gaussian energy measurements.
Here, a Gaussian energy measurement is obtained as a POVM measurement.
Then, Ref.\cite{WVT} showed that there exists a modified quantum Crooks-Tasaki work fluctuation theorem\cite{Tasaki1,Crooks} and a subsequent modified quantum Jarzynski equality for this Gaussian generalization of energy measurements.
Here, these modifications depend on the energy variance in the POVM only and are independent of the force protocol.

Second, for our purpose of treating the quantum Jarzynski equality, the initial state of the system in the quantum work protocol treated in this paper is the canonical distribution that has no quantum energy coherence.
However, in general settings, the two energy measurement approach destroys the quantum energy coherence in the initial state.
This issue has been addressed in two ways\cite{TH,Allahverdyan,SG}.

The first way is to prepare two original ensembles, one for the initial energy measurement and one for the force protocol\cite{TH}.
Then, the quantum work removes from its definition the quantum disturbance of the final state caused by the initial energy measurement.
Following the ideas of Refs.\cite{All1,All2}, Ref.\cite{Allahverdyan} characterizes this {\it untouched} work as a classical random fluctuating quantity by using a probability density function (pdf)-like function for its statistical average.
However, the weight of this pdf-like function is a quasiprobability for the eigenvalues of the initial and final Hamiltonians based on the Terletsky-Margenau-Hill distribution\cite{TMH1,TMH2} that can exhibit negative values\cite{Hartle}.
This fact limits the scope of applicability of this characterization.

The second way is to invoke von Neumann's old and original measurement protocol proposed in Ref.\cite{Neumann} (i.e., the approach of taking the partial trace of the measuring system part) and assume its validity.
In this way, for arbitrary initial states of the quantum system $S$, Ref.\cite{SG} models the protocol to measure work by a time-dependent Hamiltonian including the instantaneous von Neumann-type entangling interactions and shows that the statistics of the work performed on the quantum system $S$ exhibits non-classical correlations attributed to the initial quantum energy coherence that was destroyed by the initial energy measurement.

Finally, we comment on quantum heat engine\cite{QHE,HT,MTH}.
For us, definitions of a quantum heat engine to extract work from the quantum internal system $I$ (which usually consists of the thermodynamical system $S$ and the heat baths $\{B_m\}$) and the extracted work are the temporal control of the Hamiltonian of the system $I$, that is, the force protocol, and the loss in the expected energy of this system $I$, respectively.
However, when we consider a macroscopic external system $O$ that interacts with the system $I$ and extracts work from the system $I$, the reduced time evolution of the system $I$ is, in general, not unitary due to its openness and its interaction with the system $O$, and the validity of this definition of the extracted work is unclear.
On these points, in Ref.\cite{HT}, a novel framework is proposed for defining a quantum heat engine as the measurement process of the system $O$: (i) the system $O$ is initially in an energy eigenstate; (ii) the system $O$ interacts with the system $I$ during the force protocol; and (iii) the energy of the system $O$ is finally measured, with the extracted work defined as the difference between the final and initial energies of the system $O$.
In Ref.\cite{MTH}, using this framework, the modified quantum Jarzynski equalities are derived.

\section{Brief Account of Continuous Superselection Rules}

In the main text, we have invoked a continuous superselection rule as the concrete mechanism for the non-selective measurement.
Since this paper is primarily written for statistical physicists, we present this appendix to introduce the basic ideas and explain this mechanism in a comprehensive form.

\subsection{Basic Ideas and Motivation}
In quantum mechanics, a {\it superselection rule} refers to a selection of the observables, $\{\widehat{{\cal O}}\}$, from the Hermitian operators acting in the state space by requiring that the observables must commute with some chosen (superselection rule) Hermitian operator $\widehat{J}$:
\begin{equation}
[\widehat{{\cal O}},\widehat{J}]=0\;.\label{eq:JDef}
\end{equation}
At the same time, we decompose the state space as the direct sum (in the discrete case) or the direct integral (in the continuous case) of {\it superselection sectors} each of which is the vector space of eigenstates of $\widehat{J}$ with an eigenvalue of $\widehat{J}$.

For two state vectors $|\psi_1\ra$ and $|\psi_2\ra$ belonging to different superselection sectors (i.e., having different eigenvalues $j_1$ and $j_2$, respectively, of $\widehat{J}$), the corresponding matrix elements (i.e., their interference terms) of {\it any} observable $\widehat{{\cal O}}$ are always zero.
This fact follows immediately by applying the commutation rule for $\widehat{J}$ and $\widehat{{\cal O}}$ to the matrix elements of $\widehat{J}\widehat{{\cal O}}$.
Namely,
\begin{eqnarray}
j_1\la \psi_1|\widehat{{\cal O}}|\psi_2\ra&=&\la \psi_1|\widehat{J}\widehat{{\cal O}}|\psi_2\ra\\
&=&\la \psi_1|\widehat{{\cal O}}\widehat{J}|\psi_2\ra\\
&=&j_2\la \psi_1|\widehat{{\cal O}}|\psi_2\ra\\
&\Rightarrow&\la \psi_1|\widehat{{\cal O}}|\psi_2\ra=0\;.
\end{eqnarray}
This means that the quantum coherence between two state vectors belonging to different superselection sectors is automatically destroyed as, for an arbitrarily given superposition $|\Psi\ra=c_1|\psi_1\ra+c_2|\psi_2\ra$,
\begin{equation}
\la \Psi|\widehat{{\cal O}}|\Psi\ra=|c_1|^2\la \psi_1|\widehat{{\cal O}}|\psi_1\ra+|c_2|^2\la \psi_2|\widehat{{\cal O}}|\psi_2\ra\ \ {\rm for}\ {\rm all}\ \widehat{{\cal O}}\;,
\end{equation}
and the resultant density matrix is automatically equivalent to an exclusive mixture
\begin{equation}
\widehat{\varrho}=|c_1|^2|\psi_1\ra\la \psi_1|+|c_2|^2|\psi_2\ra\la\psi_2|\;.
\end{equation}
Within a single superselection sector, quantum coherence is not automatically destroyed.

A physical meaning of {\it continuous} superselection rule is given by a non-trivial classical observable.
Familiar examples of a continuous superselection rule are the center of mass momentum of the macroscopic measuring apparatus and the external magnetic field in the Zeeman energy of a spin.
From the definition (\ref{eq:JDef}), $\widehat{J}$ itself is an observable.
With respect to $\widehat{J}$, ideally, the quantum uncertainty principle does not hold, since $\widehat{J}$ commutes with (i.e., can be simultaneously measured with) all observables.

The motivation to invoke a continuous superselection rule\cite{Araki} is that it permits deriving a non-selective measurement (defined in Introduction) within the unitary quantum dynamics as a decoherence without using the partial trace of the measuring system part in the density matrix of the total system.
In contrast, the von Neumann theory of quantum measurement\cite{Neumann} essentially uses this partial trace to eliminate the interference terms of all observables of the measured system.
Since the information of the measuring system is completely lost in that case, we consider that the von Neumann theory unsatisfactory.

\subsection{Decoherence Mechanism}

In the following, by assuming a continuous superselection rule\cite{Araki}, we show the well-known result that, for the combined system ${S}$ of the measured system $S_0$ and a {\it macroscopic} measurement apparatus $A$ ($A$ is abstracted to a quantum system with one degree of freedom), without any interaction with the outer system of ${S}$, the off-diagonal elements of the density matrix vanish {\it dynamically} (in the situation of a non-selective measurement) in an infinite time process under the von Neumann form interaction.

The following notation and interpretations are in accord with those in Ref.\cite{Ozawa}.

We denote the state spaces of the systems $S_0$ and $A$ by ${\cal V}^{S_0}$ and ${\cal V}^A$, respectively.
We assume a measured observable $\widehat{{\mathscr{O}}}^{S_0}$ in the system $S_0$ with a discrete spectrum.
We denote the eigenvector of $\widehat{{\mathscr{O}}}^{S_0}$ with eigenvalue $x_n$ by $|x_n\ra$ and denote the initial time state vectors of the systems $S_0$ and $A$ by $|\psi\ra=\sum_n c_n|x_n\ra$ and $|\varphi\ra$, respectively.
We denote the position and momentum operators of the center of mass of the system $A$ by $\widehat{Q}^A$ and $\widehat{P}^A$, respectively.

We assume the von Neumann form interaction Hamiltonian of the system ${S}$\cite{Neumann}
\begin{equation}
\widehat{{\cal H}}_{\rm int}=-(\Lambda^A\cdot 1^A)\widehat{{\mathscr{O}}}^{S_0}\otimes \widehat{P}^A\label{eq:BB}
\end{equation}
is strong enough that we can neglect the kinetic Hamiltonian of the system $S$ by using the time parameter rescaled as $\delta t_{\rm old} \to \delta t_{\rm new}=\Lambda^A \delta t_{\rm old}$.
Here, unity $1^A$ has dimensions and $\Lambda^A$ is a dimensionless positive-valued constant.
By rescaling time as above, we reset $\Lambda^A=1$ in Eq.(\ref{eq:BB}).

Here, the center of mass momentum operator $\widehat{{P}}^A$ plays the role of a continuous superselection rule $\widehat{J}$ for the system $A$ due to the ignorability of the quantum uncertainty of $\widehat{{P}}^A$ as something like a classical observable of the macroscopic measurement apparatus $A$.
The quantum uncertainty of $\widehat{{Q}}^A$ is also ignorable.
From now on, we denote eigenvalues of the position and momentum operators of the center of mass of the measurement apparatus by $q$ and $p$, respectively.

We decompose the state space ${\cal V}^S$ of the combined system ${S}$ into the direct integral of the continuous superselection sectors
\begin{eqnarray}
{\cal V}^S&=&{\cal V}^{S_0}\otimes {\cal V}^A\\
&=&\int^\bigoplus {\cal V}^{S}(p)dp\;,
\end{eqnarray}
where we set ${\cal V}^{S}(p)\equiv {\cal V}^{S_0}$.

Then, before we apply the continuous superselection rule, the density matrix of the system ${S}$ at time $t\ge 0$ develops from $\Psi_0=|\psi\ra|\varphi\ra$ at $t=0$ to
\begin{eqnarray}
&&
|\Psi_t\ra\la \Psi_t|
\nonumber\\
&=&\sum_{m,n}c_m\bar{c}_ne^{\frac{i}{\hbar} t \widehat{{\mathscr{O}}}^{S_0}\otimes \widehat{P}^A}|x_m\ra\la x_n|
\otimes |\varphi(q)\ra\la \varphi(q)|e^{-\frac{i}{\hbar} t \widehat{{\mathscr{O}}}^{S_0}\otimes \widehat{P}^A}\\
&=&\sum_{m,n}c_m\bar{c}_n|x_m\ra\la x_n|\otimes |\varphi (q+tx_m)\ra\la \varphi(q+tx_n)|\;.
\end{eqnarray}
Here, von Neumann gave the {\it pointer position} of the measurement apparatus $A$ by the eigenvalue $q$ of $\widehat{Q}^A$\cite{Neumann}.

After we apply the continuous superselection rule, the density matrix of the system ${S}$ at time $t\ge 0$ is
\begin{eqnarray}
&&
\int^\bigoplus |\Psi_t(p)\ra\la \Psi_t(p)|dp
\nonumber\\
&=&\sum_{m,n}c_m\bar{c}_n\int^\bigoplus e^{\frac{i}{\hbar} t \widehat{{\mathscr{O}}}^{S_0}p}|x_m\ra\la x_n|e^{-\frac{i}{\hbar}t \widehat{{\mathscr{O}}}^{S_0}p}|\varphi(p)|^2dp\\
&=&\sum_{m,n}c_m\bar{c}_n\int^\bigoplus |x_m\ra\la x_n|e^{\frac{i}{\hbar}t (x_m-x_n)p}|\varphi(p)|^2dp\;,
\end{eqnarray}
where $\varphi(p)$ is the wave function of $|\varphi\ra$ in the center of mass momentum representation.
Despite the ignorability of the quantum uncertainty of $\widehat{P}^A$, the variance of the distribution $|\varphi(p)|^2$ is not zero due to the uncertainty principle when the pointer position $q$ of the measurement apparatus $A$ is initially assumed to take a definite value.

Now, we consider an arbitrarily given observable (i.e., an arbitrary polynomial of canonical variables and spin variables) of the system $S$
\begin{equation}
\widehat{{\cal X}}^S=\int^\bigoplus \widehat{{\cal X}}^S(p)dp\;,\label{eq:B8}
\end{equation}
where $\widehat{{\cal X}}^S$ is a Hermitian operator acting in ${\cal V}^S$ and $\widehat{{\cal X}}^S(p)$ is a Hermitian operator acting in ${\cal V}^S(p)$.

Due to the continuous nature of the spectrum $p$ of our superselection rule $\widehat{P}^A$, by using the Riemann-Lebesgue theorem for $x_m\neq x_n$, that is, {\it the magic trick}:
\begin{equation}
\int e^{\frac{i}{\hbar}t(x_m-x_n)p}F(p)dp\to 0\ \ {\rm as}\ \ t\to\infty\;,
\end{equation}
where $F(p)$ has finite-width, we obtain in the limit $t\to\infty$:
\begin{eqnarray}
\la{\widehat{{\cal X}}^S}\ra&=&\int \la \Psi_t(p)|\widehat{{\cal X}}^S(p)|\Psi_t(p)\ra dp\\
&\to&\sum_n |c_n|^2\int \la x_n|\widehat{{\cal X}}^S(p)|x_n\ra|\varphi(p)|^2 dp\;.
\end{eqnarray}
Namely, the interference terms in $\la{\widehat{{\cal X}}^S}\ra$ cannot be observed in the limit $t\to\infty$.
(This is because the fluctuations of $\la{\widehat{{\cal X}}^S}\ra$ less than the uncertainty width of ${\widehat{{\cal X}}^S}$ cannot be observed.)
This holds in the case of arbitrary $\widehat{{\cal X}}^S$.
Then, we obtain the resultant density matrix in the limit $t\to\infty$
\begin{equation}
\int^\bigoplus |\Psi_t(p)\ra\la \Psi_t(p)|dp\to \sum_n |c_n|^2\int^\bigoplus |x_n\ra\la x_n||\varphi(p)|^2 dp\;.
\end{equation}
This is nothing but the state after the non-selective measurement in the combined system ${S}=S_0+A$.
Here, the ignorable but finite uncertainty of $\widehat{P}^A$ causes this result via the interaction $\widehat{{\cal H}}_{\rm int}$ between the systems $S_0$ and $A$ during an infinite time range.

\subsection{Remark on a Circumvented No-go Theorem}

As an important remark, if we do not assume a continuous superselection rule, due to the no-go theorem proved by Wigner, Fine, Shimony and Araki\cite{Araki2,NGthm1,NGthm2,NGthm3}, in the combined quantum system $S_0+A$ the quantum coherence cannot vanish by a unitary time-dependent process within either a finite time-interval\cite{NGthm1,NGthm2,NGthm3} or an infinite time-interval\cite{Araki2}.
This no-go theorem is circumvented by the above mechanism of non-selective measurement in the system $S_0+A$.
However, due to the {\it unitarity} of time evolution, to work this mechanism, we require an infinite time range process from the microscopic point of view.

\end{appendix}

\end{document}